\documentclass[smallextended]{svjour3}       
\usepackage{pgfplots}
\usepackage{tikz-3dplot}
\usepackage{soul}
\tdplotsetmaincoords{60}{135}
\pgfplotsset{compat=newest}
\usepackage{graphicx}
\usepackage{tikz}
\usetikzlibrary{calc,patterns,decorations.pathmorphing,decorations.markings, quotes,angles}
\usepackage{tikz-3dplot}

\makeatletter
\tikzset{
    dot diameter/.store in=\dot@diameter,
    dot diameter=3pt,
    dot spacing/.store in=\dot@spacing,
    dot spacing=10pt,
    dots/.style={
        line width=\dot@diameter,
        line cap=round,

        dash pattern=on 0pt off \dot@spacing
    }
}
\makeatother

\smartqed  
\usepackage[latin1]{inputenc}
\usepackage{graphicx}
\usepackage{tikz}
\usetikzlibrary{calc,patterns,decorations.pathmorphing,decorations.markings, quotes,angles}
\usepackage{tikz-3dplot}
\usepackage{amsmath,amsfonts,amssymb,amscd}
\usepackage{wasysym}
\usepackage{wrapfig}
\usepackage[colorlinks,linkcolor=blue,citecolor=blue,urlcolor=blue]{hyperref}
\usepackage{natbib}
\setcitestyle{round}

\usepackage{tikz}
\usetikzlibrary{calc,patterns,decorations.pathmorphing,decorations.markings}

\newcommand{\erm}{{\rm e} }

\usepackage[graphicx]{realboxes}

\newcommand{\Id}{{\bf 1}  }

\newcommand{\K}{\mathrm{K}}

\newcommand{\R}{\mathbb{R}}

\newcommand{\Oc}{\mathcal{O}}

\newcommand{\sg}{\sigma}
\newcommand{\gm}{\gamma}

\newcommand{\ep}{\epsilon}

\newcommand{\mat}[1]{\mathbf{#1}}

\newcommand{\ov}{\overline}

\newcommand{\Io}{ {\,\rm I}_\circ}

\def\Z{{\mathbb Z}}

\def\R{{\mathbb R}}

\newcommand{\nd} {\noindent}

\begin{document}

\setstcolor{red}

\title{Spin-Orbit Synchronization  and Singular Perturbation Theory}



\author{    Clodoaldo Ragazzo    \and
           Lucas Ruiz dos Santos 
}

\institute{C. Ragazzo  (ORCID 0000-0002-4277-4173)\at
 Instituto de Matem\'{a}tica e Estat\'{i}stica, Universidade de S\~{a}o Paulo, 05508-090 S\~{a}o Paulo, SP, Brazil
 \\ \email{ragazzo@usp.br}
\and
L.S. Ruiz   (ORCID 0000-0002-5705-5278) \at
Instituto de Matem\'{a}tica e Computa\c{c}\~{a}o, Universidade Federal de Itajub\'{a}, 37500-903 Itajub\'{a}, MG, Brazil
\\ \email{lucasruiz@unifei.edu.br}
}

\date{Received: date / Accepted: date}

\maketitle

\begin{abstract}
  In this study, we formulate a set of differential equations for a binary system to describe the secular-tidal evolution of orbital elements, rotational dynamics, and deformation (flattening), under the assumption that one body remains spherical while the other is slightly aspherical throughout the analysis. By applying singular perturbation theory, we analyze the dynamics of both the original and secular equations. Our findings indicate that the secular equations serve as a robust approximation for the entire system,
  often representing a slow-fast dynamical system.
 Additionally, we explore the geometric aspects of spin-orbit resonance capture,
 interpreting it as a manifestation of relaxation oscillations within singularly perturbed systems.
\end{abstract}

\vskip 1.0 truecm

\nd { \bf \large Preamble}

\vskip 0.5 truecm

\nd This work is dedicated to the memory of Prof. Jorge Sotomayor, a teacher and friend. Unlike typical mathematical publications, this paper contains no theorems. Instead, it focuses on applications of methods in Ordinary Differential Equations (ODE), a field where, as CR heard from Prof. J. K. Hale, ``techniques such as averaging, normal forms, and challenges like the N-body problem, Hilbert's XVI problem, and the Lorenz equation, become crucial in research, overshadowing the established general theory.''

CR had the honor of collaborating with Prof. Sotomayor for nearly two decades at the Instituto de Matem\'atica e Estat\'\i stica da Universidade de S\~ao Paulo, where our daily interactions were enriched by his humorous insights on life. More than just a brilliant mathematician, he was vivacious, joyful, and optimistic. He often shared a belief  that ``for a mathematical field to flourish, it must engage with other sciences or mathematical areas''. Prof. Sotomayor's work in ODEs, a discipline rooted in Isaac Newton's efforts to solve physical and geometrical problems, significantly advanced both the theoretical aspects of ODEs through his studies on bifurcations and their practical applications, notably in differential geometry's lines of curvature.

His students and friends hope that his  legacy endures: to approach ODE with joy and happiness.

\section{Introduction}
\label{intro}

The foundations of differential equations trace back to Newton's pioneering work in mechanics and differential calculus.
Newton  grounded the law of gravitation mathematically and   solved  the equations for the
motion of two  bodies. However, the Newtonian model primarily considers celestial bodies as
point masses, a simplification that has its limitations given that celestial entities
have finite dimensions.

Planets and substantial satellites  exhibit a near-spherical shape.
Despite being relatively minuscule compared to their respective diameters,
the deformations induced by spin and tidal forces have a considerable impact, instigating significant alterations in both
rotation rates and orbits. It is worth noting that all the major satellites within our solar system, including the Moon, operate in a 1:1 spin-orbit resonance
(see, e.g., \cite{Murray}), they complete a single rotation on their axis for every orbit around the  planet. Mercury, however, maintains a 3:2 spin-orbit resonance, undergoing three rotations on its axis for every two revolutions around the Sun. Furthermore, a majority of these celestial entities follows elliptical orbits characterized by low eccentricity.
Deciphering how this dynamic state was attained, along with determining the associated time scales,
holds substantial significance in the scientific realm.

The goal of this study is to introduce  equations to describe  the perturbative impact of
deformations on the motion of two spherical bodies influenced by gravitational interaction.
Subsequently, we demonstrate that in certain limiting scenarios, which bear physical relevance,
these equations can be analyzed using the mathematical apparatus of singular perturbations.

The earliest and most basic deformation model accounting for energy dissipation was put forth by George Darwin \cite{darwin1879bodily}, son of the renowned biologist Charles Darwin. Darwin built upon previous studies \cite{thomson1863xxvii} concerning the deformation of an elastic, homogeneous, incompressible sphere, extending the results to address a body constituted of a homogeneous, incompressible, viscous fluid.

Subsequent to Darwin, a significant advancement came with the introduction of Love numbers \cite{love1911}.
When the tidal force is decomposed in time via its Fourier components and in space through  spherical-harmonic  components,
the Love number for a specific harmonic frequency and spherical-harmonic mode is a  scalar that
correlates the amplitude of the tidal force to the deformation's amplitude.
Essentially, Love numbers act as functions within the frequency space, offering a phenomenological approach to elucidate force-deformation relationships. Estimates of Love numbers can be derived from observational data.

Over the past 70 years, there has been a prolific output of scientific literature focusing on the tidal effects on the motion of celestial bodies. While it is  challenging to encompass the breadth of these studies, we will mention a few we are particularly acquainted with.

Kaula  \cite{kaula1964tidal}  evaluated the rate of change of the orbital elements using Love numbers for each harmonic mode (see \cite{boue2019tidal}  and \cite{efroimsky2012bodily}  for
further insights on the work of Kaula).
Numerous other scholars  have investigated equations accounting for deformations averaged over orbital motion. Some important works in this area are: \cite{goldreich1966final}, \cite{singer1968origin}, \cite{alexander1973weak}, and \cite{mignard1979evolution} (low-viscosity scenarios); and  \cite{makarov2013no}, \cite{ferraz2013tidal}, \cite{correia2014deformation}, \cite{ferraz2015tidal}, and \cite{boue2016complete}, \cite{folonier2018tidal}, \cite{ferraz2019planetary}, \cite{ferraz2020tidal},
\cite{ferraz2021tides}
(low and  high-viscosity scenarios).

In this paper, for simplicity while maintaining physical relevance, we make the following assumptions:
\begin{itemize}
\item[$1)$] The first body is deformable, nearly spherical at all times; 
\item[$2)$] The second body, which is the tide-raising body, is a point mass; 
\item[$3)$] The spin (or rotation vector) of the deformable body remains perpendicular to the orbital plane.
  \end{itemize}
  The foundational equations for the orbit and rotation of the extended body are standard.
  Various equations exist in the literature detailing the deformation of extended bodies.
  We utilize the equations provided in \cite{rr2017}, without the term accounting for the inertia of deformations
  \cite{correia2018effects}.

  The reduced and averaged equations we introduce here are not  novel. Excluding centrifugal deformations,
  they match those in \cite{correia2022tidal}. Our analysis parallels the approach in \cite{correia2014deformation},
  Section 5. The primary contributions of this paper include: 
  \begin{itemize}
    \item[$1)$]Clearly stating mathematical assumptions used in deriving the averaged and reduced equations; 
 \item[$2)$] Framing the averaged equations as a slow-fast system; 
 \item[$3)$] Beginning a geometric examination of the slow system using numerically generated figures to
   illustrate the ``relaxation jumps''.
   \end{itemize}
   We adopt the geometric method set out by Fenichel \cite{fenichel1971persistence},
   \cite{fenichel1974asymptotic},
   \cite{fenichel1977asymptotic}, \cite{fenichel1979geometric}, and \cite{krupa2001relaxation}
   without fully verifying all the assumptions.
   A comprehensive mathematical analysis of the equations presented may necessitate extensive research.

The paper is structured as follows:

In Section \ref{eqsec}, we outline the core equations of the system. We assess the magnitude of various terms and introduce a parameter representing the minor nature of the deformations.

In  Section \ref{avsec1}, we examine the limit when deformations approach zero, averaging them over orbital motion. This leads to equations with ``passive  deformations'' that do not  influence the orbit.

In  Section \ref{avsec2}, we suggest that for minor deformations, the primary equations possess an attracting invariant manifold matching the deformations from Section \ref{avsec1}. This manifold's existence depends  on the body's rheology.
As the body becomes more viscous, the manifold becomes less attractive \footnote{This counterintuitive claim is associated with the omission of deformation inertia. In the equation for the damped harmonic oscillator $m \ddot x = -x - \eta \dot x$, the solutions converge to zero more rapidly as the damping coefficient $\eta$ increases. If the inertia coefficient is zero, the equation simplifies to $\eta \dot x = -x$, leading to the opposite effect: $x(t) = e^{-t/\eta} x(0)$.
}.
Given the enhanced spin-orbit coupling at high viscosity, assessing the credibility of our calculations and assumptions in this section presents a compelling mathematical challenge.

In  Section \ref{avesec3}, we average the orbital and spin equations based on the preceding section's invariant manifold.

Section \ref{singular} reveals that
 the averaged equations exhibit a slow-fast split. The fast
 variable is the body's spin, while the slower variables are orbital eccentricity and the semi-major axis.

 In Section \ref{secso}, we delineate a condition for the folding of the slow manifold and provide a numerical illustration of its geometry. We also present a geometric interpretation of the dynamics within this manifold, emphasizing rapid spin transitions as instances of ``relaxation jumps''  \cite{mishchenko2013differential}, \cite{krupa2001relaxation}.

Section \ref{conclusions} concludes the paper, recapping the pivotal mathematical queries
regarding the simplification of the initial equations and the dynamics of the reduced equations.

This paper was written concurrently with a companion paper \cite{rr2024b}, which has a more physics-oriented content. The focus of \cite{rr2024b} is on the implications for dynamics of using rheological models more complex than the one employed here.

\section{The fundamental equations.}

\label{eqsec}

Let $m_0$ and $m$ represent the masses of two celestial bodies, which could be a planet and a star, or a planet and a satellite, etc. The body with mass $m_0$ is treated as a point mass, while the body with mass $m$ is always a small deformation of a spherical body with a moment of inertia $\Io$. We assume that the deformations do not alter the volume of the body, implying that $\Io$ remains constant, a result attributed to Darwin
\cite{rochester1974changes}. Often, we will refer to the bodies simply as the point mass and the body.

For convenience, we write the deviatoric part of the moment of inertia matrix $\mathbf{I}$ in non-dimensional form:
\begin{equation}
  \mathbf{I}= \Io \big(\Id-\mathbf{b}\big)
\end{equation}
where $\Id$ is the identity  and $\mathbf{b}$ is a symmetric and traceless matrix. We denote matrices and vectors in bold face. The matrix $\mat b$ is termed the deformation matrix.

Consider an orthonormal frame $\{\mat e_1,\mat e_2,\mat e_3\}$. We assume that the vector $\mathbf x$, from the center of mass of the body to the point mass, lies in the plane spanned by $\{\mat e_1,\mat e_2\}$. The angular velocity of the body, $\boldsymbol \omega$, is perpendicular to the orbital plane, represented as $\boldsymbol \omega=\omega \mat e_3$. The deformation matrix is given by:
\begin{equation}
 \mat b=\left(
\begin{array}{ccc}
 b_{11} &  b_{12} & 0 \\
  b_{12} &  b_{22} & 0 \\
 0 & 0 &  b_{33} \\
\end{array}
\right)\,,\quad \text{with}\quad b_{33}=-b_{11}-b_{22}\,.
\label{B}
\end{equation}

Under the given assumptions, Newton's equation for the relative position is expressed as:
\begin{equation}  
  \ddot {\mathbf x} = G(m_0+m)  \bigg\{
- \frac{\mathbf x}{|\mathbf x|^3}
+\frac{\Io}{m}\bigg(-\frac{15}{2}\frac{1}{|x|^7} 
(\mathbf b \mathbf x \cdot\mathbf x)\mathbf x  
+3\frac{1}{|\mathbf x|^5}    \mathbf b\mathbf x\bigg)
\bigg\}\,,
\label{xeq}
\end{equation}
where it is assumed that in the region occupied by the body, the gravitational field of the point mass is accurately represented by its quadrupolar approximation.

The spin angular momentum of the body is denoted by $ \boldsymbol{\ell}_s=\ell_s\mat e_3$, with the index $s$ representing spin, and is defined as:
\begin{equation} \ell_s=\omega \Io(1-b_{3 3})\,.\label{eldef}
\end{equation}

In the context of the quadrupolar approximation, Euler's equation for the variation of $\ell_s$ is: 
\begin{equation} \label{leq}
  \dot\ell_s=
  - \frac{3G \Io m_0}{\|\mat x\|^5}\Big\{
x_1x_2 (b_{ 22}-b_{ 11})+b_{12} \left(x_1^2-x_2^2\right)\Big\}.
\end{equation}  

For a rigid body, a specific frame exists, known as the body frame, in which the body remains stationary and its angular momentum with respect to this frame is zero. Similarly, for a deformable body, there is an equivalent frame, called the Tisserand frame, where the body's angular momentum is null.
  The orientation of
  the Tisserand frame $\K:= \{\mat e_{T1}, \mat e_{T2}, \mat e_{T3}\}$ with respect to the inertial frame $\kappa := \{\mat e_{1}, \mat e_{2}, \mat e_{3}\}$ is given by
\begin{equation}
\mat R(\phi) = \left( \begin{array}{ccc} 
 \cos \phi  & -\sin \phi & 0 \\ 
 \sin \phi  & \cos \phi & 0 \\
 0 & 0 & 1
\end{array}  \right):\K\to \kappa
\label{rot}
\end{equation}
and by definition, the rate of change of the angle $\phi$ is given by:
\begin{equation}
  \dot \phi=\omega\,.
\end{equation}

  To complete the set of equations (\ref{xeq}) and (\ref{leq}), we require additional equations for the deformation matrices. These equations were derived within the Lagrangian formalism and utilizing what was termed the  ``Association Principle,'' as detailed in \cite{rr2015}, \cite{rr2017} (see, also \cite{gev2020} addressing the treatment of Andrade rheology, \cite{ragazzo2022librations} extending to bodies with permanent deformation, and \cite{gev2021} and \cite{gevorgyan2023equivalence} exploring the relations with the rheology of layered bodies).

 To maintain simplicity in mathematical expressions, we consider only the basic rheology of ``Kelvin-Voigt'' combined with self-gravity here. The exploration of more generalized rheologies,
 which might introduce new time scales to the problem, is reserved for a companion paper \cite{rr2024b}.

The Tisserand frame of the body is the natural frame to present the equations for deformations. In this frame, the deformation matrix and the position vector are denoted by capital letters as follows:
\begin{equation}
   \mathbf B=\mat R(\phi)\mat b\mat R^{-1}(\phi)\qquad \mat X= \mat R^{-1}(\phi)\mat x \,.
\end{equation}
The governing equation for $\mat B$ is:
\begin{equation}
 \eta\dot{\mathbf{B}} + (\gm+\alpha)\mat B= \mathbf{F}\,,
 \label{DFM}
\end{equation}
where:
\begin{itemize}
\item[$\bullet$] $\gm$, with dimensions of $1/$time$^2$, is a parameter representing the self-gravity rigidity of the body; a larger $\gamma$ indicates a stronger gravitational force holding the body together.
\item[$\bullet$] $\alpha$, also with dimensions of $1/$time$^2$, signifies the elastic rigidity of the body; for a fluid body, $\alpha=0$.
\item[$\bullet$] $\eta$, dimensions of  $1/$time, is a viscosity parameter; a body with a larger $\eta$ is harder to deform at a given rate compared to a body with a smaller $\eta$.
\item[$\bullet$] $\mathbf{F}$, with dimensions 1/time$^2$, is the force matrix in the Tisserand frame $\K$:
\begin{equation}\renewcommand{\arraystretch}{1.5}
  \begin{array}{l l l}
    \mathbf{F}&:=\mat C+\mat S \quad&\text{Deformation force}\\ & & \\
    \mat C&:= \frac{\omega^2}{3}\left(\begin{matrix}
1& 0 & \ \ 0 \\ 0 & 1 & \ \ 0
\\ 0 & 0 & -2
\end{matrix}  \right)\quad&\text{centrifugal force}\\ & & \\
     \mat S&:=
                   \frac{3G m_0}{|\mat X|^5}\left(\mathbf{X}\otimes\mathbf{X}-
             \frac{|\mat X|^2}{3}\Id\right)\quad& \text{Tidal force}
\end{array}
\label{F}
\end{equation}
\end{itemize}
where $\mat X\otimes\mat X$ is a matrix with entries
$\big(\mat X\otimes\mat X\big)_{ij}=X_iX_j$.

To determine the Love number function associated with the deformation equation (\ref{DFM}), we consider a simple harmonic force term of the form
\[ \mat F(t)=\mat {\widehat F}\erm^{\sigma t}\]
where $\mat {\widehat F}$ is a complex amplitude matrix, and $\sigma\in\R$ is the constant forcing frequency. Assuming a solution of the form $\mat B(t)=\mat {\widehat B}\erm^{\sigma t}$, we derive the relationship between the complex amplitudes as
\begin{equation}
     \mat {\widehat B}=\underbrace{\frac{1}{\gamma+\alpha+i\, \eta \sigma}}_{C(\sigma)}
     \mat {\widehat F}=
     \left(\frac{1}{\gamma+\alpha}\right)\frac{1}{1+i\, \tau \sigma}\mat {\widehat F}=
     \left(\frac{1}{\gamma+\alpha}\right)\frac{1-i\, \tau \sigma}{1+ \tau^2 \sigma^2}\mat {\widehat F}
\end{equation}
where $C(\sigma)$ is the complex compliance and 
\begin{equation}
     \tau:=\frac{\eta}{\gamma+\alpha}\quad\text{represents the time constant.}
     \label{deftau}
\end{equation}

The complex Love number $k_2(\sigma)$, commonly defined differently (see, e.g., \cite{rr2017}), is proportional to the complex compliance $C(\sigma)$ as outlined in \cite{mathews2002modeling} (paragraph 21):
\begin{equation}
     k_2(\sigma)=\frac{3G\Io}{R^5}C(\sigma)=
    \left(\frac{3G\Io}{R^5}\frac{1}{\gamma+\alpha}\right)
    \frac{1-i\, \tau \sigma}{1+ \tau^2 \sigma^2}=k_\circ
     \frac{1-i\, \tau \sigma}{1+ \tau^2 \sigma^2}\,,\label{lov}
\end{equation}
where the number $k_\circ:=\frac{3G\Io}{R^5}\frac{1}{\gamma+\alpha}$ denotes the secular Love number, representing the value of $k_2(\sigma)$ for static forces ($\sigma=0$).

In the case of a fluid body, the elastic modulus $\alpha$ is zero, and
\begin{equation} k_{\circ}=k_f:=\frac{3G\Io}{R^5}\frac{1}{\gamma}\quad\text{fluid Love number.}
  \label{flove}
\end{equation}
 The body is held together solely by self-gravity. For a homogeneous fluid body of any density, $k_f=3/2$. As discussed in \cite{ragazzo2020theory}, this represents the maximum possible value of $k_f$ when the density of the body increases towards the center. Given that for any non-null elastic rigidity $\alpha>0$, $k_f>k_\circ$, we conclude that for any stably stratified body,
\begin{equation}
  k_\circ=\frac{3G\Io}{R^5}\frac{1}{\gamma+\alpha}\le \frac{3}{2}. \label{kcirc}
\end{equation}

{\it Historical note.} Darwin was the pioneer in deriving equation  (\ref{lov}), while examining tides on a
homogeneous body composed of viscous fluid. In page 13 of \cite{darwin1879bodily}, Darwin stated:
``Thus we see that the tides of the viscous sphere are the equilibrium tides of a fluid sphere as $\cos\epsilon : 1$, and that there is a retardation time $\frac{\epsilon}{\sigma}$''. In his paper, $\nu$ denotes fluid viscosity, and $\tan \epsilon= \frac{19}{2}\frac{\nu }{g R \rho} \sigma$, where $g$ represents surface gravity, and $\rho$ is the mass per unit volume of the body.

Given that for a homogeneous fluid body $k_\circ=k_f=3/2$, Darwin's statement can be reformulated as
\begin{equation}\begin{split}
& k_2=\frac{3}{2}\cos\ep \, \erm^{-i \ep}= \frac{3}{2}\frac{1}{\sqrt{1+\tan^2\ep}}\erm^{-i \ep}=
\frac{3}{2}\frac{1}{\sqrt{1+\tau^2\sigma^2}}\erm^{-i \ep}= \frac{3}{2}\frac{1}{1+i\tau\sigma}
\,,\\ & \text{where}\\
& \tan\ep=\tau\sigma \quad\text{and} \quad
\tau=\frac{19}{2}\frac{\nu}{g R\rho}\,.\end{split}\label{k2Darwin}
\end{equation}

Utilizing the relationships for a homogeneous spherical body, $\Io=\frac{2}{5}m R^2$, $g=\frac{Gm}{R^2}$, and $\rho=m/\frac{4\pi R^3}{3}$, where $m$ is the mass and $R$ is the radius of the fluid body, and from the relations $ k_{\circ}=k_f=\frac{3}{2}=\frac{3\Io G}{R^5}\frac{1}{\gm}$ and $ \tau=\frac{\eta}{\gm}=\frac{19}{2}\frac{\nu}{g R\rho}$, we deduce
\begin{equation}
\eta=\frac{152\pi}{15}\frac{R}{m}\nu \,,\label{etanu}
\end{equation}
which aligns with a relation in \cite[Eq. (39)]{correia2018effects}.

The theory developed by Darwin \cite{darwin1879bodily}, \cite{darwin1880secular} has predominantly been applied in the frequency domain. Influenced by Darwin's work, Ferraz-Mello \cite{ferraz2013tidal} formulated an equation for the motion of the surface of the body under tidal forcing in the time domain. When $\alpha=0$, the model in \cite{correia2014deformation} with $\tau_e=0$, the model in \cite{ferraz2013tidal}, and equation (\ref{DFM}) are all equivalent (our $\tau$ corresponds to the $\tau$ in \cite{correia2014deformation}, which is equal to  the parameter ``$1/\gamma$'' used in \cite{ferraz2013tidal}). See \cite{correia2014deformation}, paragraph above equation (90), and \cite{ferraz2015small} for the equivalence between the models in \cite{ferraz2013tidal} and \cite{correia2014deformation}.

\section{Zero deformation limit.}
\label{avsec1}

In numerous celestial mechanics problems, bodies maintain near-spherical shapes at all times, which can be reformulated as
\begin{equation}
\|\mathbf{B}\|\ll 1\,,\quad\text{where}\quad \|\mathbf{B}\|^2=\frac{1}{2}\sum_{ij}B_{ij}^2 \,.
\end{equation}
Given that equation (\ref{DFM}) for $\mat B$ is linear, $\|\mat B\|$ is small if, and only if, $\|\mat F\|$ is small.

The relative motion between two nearly spherical bodies approximates Keplerian motion. Let $a$, $n$, and $e$ represent the semi-major axis, the mean motion (period/(2$\pi$)), and the eccentricity of the Keplerian ellipses, respectively. The magnitude of the force terms in the deformation equation (\ref{DFM}) is proportional to the following characteristic frequencies:
\begin{equation}\begin{split}
\mat S &= \frac{3G m_0}{|\mat x|^5}\left(\mathbf{x}\otimes\mathbf{x}-
\frac{|\mat x|^2}{3}\Id\right)\approx\frac{Gm_0}{a^3}= \frac{m_0}{m+m_0} n^2 \quad \text{tidal force}\,;\\
\mat C&= -\left(\boldsymbol{ \omega}_{\alpha}\otimes\boldsymbol{\omega}_{\alpha} -
\frac{\|\boldsymbol \omega_{\alpha}\|^2}{3}\Id\right)\approx 2  \omega^2\quad\text{centrifugal force}\,.
\end{split}\label{ordF}
\end{equation}

The forces on the right-hand side of equation (\ref{DFM}) are counteracted by the body's self-gravity and possibly elastic rigidity $\alpha\ge 0$.
  The static deformations are then given by \[
    \mat B=\frac{\mat C}{\gamma+\alpha}+\frac{\mat S}{\gamma+\alpha}=k_\circ\frac{R^5}{3G\Io}
    \big(\mat C+\mat S\big)\,,\]
where we used $k_\circ:=\frac{3G\Io}{R^5}\frac{1}{\gamma+\alpha}$.

The order of magnitudes in equation (\ref{ordF})  and inequality \eqref{kcirc} imply
\begin{equation}
  \|\mat B\|\le \frac{R^5\omega^2}{G\Io}+\frac{m_0R^5}{2a^3\Io}. 
\end{equation}
This indicates that the region in phase space
defined by the following inequalities:
\begin{equation}
  \zeta_c:= \frac{R^5 \omega^2}{G\Io}\ll 1
  \quad \text{and} \quad \zeta_{\scriptscriptstyle T} := \frac{m_0 R^5}{2\Io a^3}
\ll 1
\label{zeta}
\end{equation}
adheres to the small deformation hypothesis.

\subsection{The Zero Deformation Limit}

Define the compliance $\epsilon_d$, where $d$ denotes deformation, as follows:
\begin{equation}
  \epsilon_d:=\frac{1}{\gamma+\alpha}\quad \text{dimension of time}^2\,.
\end{equation}
We  then express
\begin{equation}
\mathbf B=\epsilon_d \widetilde{\mathbf B}
\end{equation}
and substitute into equations (\ref{xeq}), (\ref{eldef}), (\ref{leq}), and (\ref{DFM}) to yield
\begin{equation}
\begin{split}  
  \ddot {\mathbf x} &= G(m_0+m)  \bigg\{
- \frac{\mathbf x}{|\mathbf x|^3}
+\epsilon_d\frac{\Io}{m}\bigg(-\frac{15}{2}\frac{1}{|x|^7} 
(\widetilde{\mathbf b} \mathbf x \cdot\mathbf x)\mathbf x  
+3\frac{1}{|\mathbf x|^5}    \widetilde{\mathbf b}\mathbf x\bigg)
\bigg\}\\
\dot\ell_s&=
- \epsilon_d\frac{3G \Io m_0}{\|\mat x\|^5}\Big\{
x_1x_2 (\tilde b_{ 22}-\tilde b_{ 11})+\tilde b_{12} \left(x_1^2-x_2^2\right)\Big\}\\
\ell_s&=\omega \Io(1-\epsilon_d \tilde b_{3 3})\\
\tau\dot{\widetilde{\mathbf{B}}} &+ \widetilde{\mat B}= \mathbf F
\end{split}\label{epd}
\end{equation}
where $\tau$ is defined in \eqref{deftau} and
  $\tilde{\mat b}=\mat R(\phi)\widetilde{\mathbf B}\mat R^{-1}(\phi)$.

The zero deformation limit is defined by:
\begin{equation}
\epsilon_d=\frac{1}{\gamma+\alpha}\rightarrow 0 \quad\text{while}\quad
\tau=\frac{\eta}{\alpha+\gamma}\ \text{remains constant}\,.
\end{equation}

 In the zero deformation limit, equation (\ref{epd}) simplifies to:
\begin{equation}
    \begin{split}  
  \ddot {\mathbf x} &=- G(m_0+m)  \frac{\mathbf x}{|\mathbf x|^3}\\
 \dot\ell_s&=\dot \omega \Io=0\\
\tau\dot{\widetilde{\mathbf{B}}} &+ \widetilde{\mat B}=\mathbf F
\end{split}\label{epd2}
\end{equation}
In this scenario, the body spin,  $\omega$, remains constant and $\mat x$ follows a Keplerian ellipse.

To describe the Keplerian orbits, we  change  from variables $(\mat x,\dot{\mat x})$ to $\ell\in \R$ (orbital angular momentum), $\mat A$ (the Laplace vector), and $f$ (the true anomaly), defined as:
\begin{equation}
\renewcommand{\arraystretch}{1.5}
\begin{array}{rll}
\ell \mat e_3&= \boldsymbol{\ell}=\mu \mat  x \times\dot{\mat x} &\text{orbital angular momentum}\\
\mat A&= \frac{1}{c}\dot{\mat x}\times \boldsymbol{\ell}-\frac{\mat x}{|\mat x|}&\text{Laplace vector}
\end{array}\label{laplace}
\end{equation}
where
\begin{equation}
  \mu=\frac{m_0 m}{m_0+m}=\text{reduced mass}\,,\quad c=G m m_0\,.
  \label{reduced}
\end{equation}
The Laplace vector is normalized such that $\|\mat A\|=e$ is the orbital eccentricity
and it points towards the periapsis, where $\|\mat x\|$ is minimized.

The three vectors 
\begin{equation}
  \mat e_A:= \frac{\mat A}{|\mat A|}\, ,\qquad\mat e_H:= \mat e_3 \times  \mat e_A \,,\qquad\mat e_3
  \label{orb}
\end{equation}
constitute an orthonormal basis, expressed in terms of the inertial frame basis vectors as
\begin{equation}
  \mat e_A:= \cos\varpi \mat e_1+\sin\varpi \mat e_2\, ,\qquad
  \mat e_H:= - \sin\varpi \mat e_1+ \cos\varpi \mat e_2\,.
\label{orbp}
\end{equation}
Here, $\varpi$ denotes the longitude of the periapsis, the angle between $\mat e_A$ and $\mat e_1$.

The orbit is represented by 
\begin{equation}
\begin{split}
    \mat x&=  r \mat R(f+\varpi)\mat e_1=r(\cos (f+\varpi)\,  \mat e_1+\sin (f+\varpi)\, \mat e_2)\\
    &=r(\cos f\,  \mat e_A+\sin f\, \mat e_H)\,,
\end{split}
\label{kepp}
\end{equation}
where $\mat R$ is the rotation matrix about the axis $\mat e_3$, as given in equation (\ref{rot}), and
$r(t)=\|\mat x(t)\|$.

\subsection{Passive deformations.}

The equations at the zero deformation limit (\ref{epd2}) in the new variables become
(see, e.g., \cite{Murray} for  details):
\begin{equation}
    \begin{split}  
  \dot {\mat A} &=0
  \\
  \dot \ell&=0
  \\ \dot f&=\frac{\mu \ell}{r^2}\,,\ \  \text{where}\  \ r=\frac{a(1-e^2)}{1+e \cos f}=
  \frac{\ell^2}{\mu c}\frac{1}{1+e \cos f}\\
\dot \omega&=0\\
\tau\dot{\widetilde{\mathbf{B}}} + \widetilde{\mat B}&= \mathbf{C}+\mathbf S
\end{split}\label{epd3}
\end{equation}
where $\mathbf C$ and $\mathbf S$ are given in equation (\ref{F}).

In order to write $\mat S$ in a convenient way, we define the matrices
 {\small
  \begin{equation}
    \mathbf{Y}_{-2}:=
\frac{1}{\sqrt{2}}\left(\begin{matrix}
1 & i & 0 \\ i & -1 & 0
\\ 0 & 0 & 0
\end{matrix}  \right)\quad
\mathbf{Y}_0:=
\frac{1}{\sqrt{3}}\left(\begin{matrix}
1& 0 & \ \ 0 \\ 0 & 1 & \ \ 0
\\ 0 & 0 & -2
\end{matrix}  \right)\quad
\mathbf{Y}_2:=
\frac{1}{\sqrt{2}}\left(\begin{matrix}
1 & -i & 0 \\ -i & -1 & 0
\\ 0 & 0 & 0
\end{matrix}  \right),
\label{Yj}
\end{equation}}
with $\mat Y_{-2}=\ov{\mat Y}_2$, where the overline  represents
complex conjugation.
These matrices have a simple transformation rule with respect to  rotations about the axis
$\mat e_3$, namely
  \begin{equation}
    \mat R(\theta)\mat Y_j\mat R^{-1}(\theta)=\erm^{i\, j\, \theta} \,\mat Y_j\,,\quad j=-2,0,2\,.
\label{eig}    \end{equation}

Using 
\begin{equation}\begin{split}
    \mathbf X&=\mat R^{-1}(\phi)\mat x=  r \mat R(f+\varpi-\phi)\mat e_1\\
    &=
    r(\cos (f+\varpi-\phi)\,  \mat e_1+\sin (f+\varpi-\phi)\, \mat e_2)\, ,
    \end{split}
\end{equation}
the tidal-force matrix in equation (\ref{F}) can be written as  
\begin{equation}
   \mat S =\frac{3G m_0}{r^3}\mat R(f+\varpi-\phi)\left(\mathbf{e}_1\otimes\mathbf{e_1}-
                     \frac{1}{3}\Id\right)\mat R^{-1}(f+\varpi-\phi).
                 \end{equation}
In the basis $\{\mat Y_{-2},\mat Y_0,\mat Y_2\}$ 
\begin{equation}
  \mat e_1\otimes\mat e_1-
  \frac{1}{3}\Id =\frac{1}{2}\bigg\{\frac{\mat Y_{-2}}{\sqrt 2}+\frac{\mat Y_0}{\sqrt 3}+
  \frac{\mat Y_2}{\sqrt 2}\bigg\}
\end{equation}
that implies
\begin{equation}\begin{split}
  \mat S&=\frac{3 G m_0}{r^3} \mat R_3(f+\varpi-\phi)\Big\{\mat e_1\otimes\mat e_1-  \frac{1}{3}\Id\Big\} \mat R^{-1}_3(f+\varpi-\phi)\\ & =\frac{3 G m_0}{2r^3}
 \bigg\{\erm^{-2 i (f+\varpi-\phi)}\frac{ \mat Y_{-2}}{\sqrt 2}+\frac{\mat Y_0}{\sqrt 3}+
  \erm^{2 i (f+\varpi-\phi)}\frac{\mat Y_2}{\sqrt 2}\bigg\}.\end{split}
  \label{tf}
  \end{equation}

 In equation (\ref{tf}), the variables $r$, $f$, and $\phi=\omega t$ are dependent on $t$.

 To solve the equation $\tau\dot{\widetilde{\mathbf{B}}} + \widetilde{\mat B}= \mathbf{C}+\mathbf S$,
 we do a harmonic analysis of the tidal force in equation (\ref{tf}) using:
\begin{equation}
 \left(\frac{r}{a}\right)^{n^\prime}\erm^{i m f}=\sum_{k=-\infty}^\infty X^{n^\prime,m}_k(e) \erm^{i k M}\,,
 \label{hanseneq}
 \end{equation}
where $M$ denotes the mean anomaly, $\dot M=n$, and $X^{n^\prime,m}_k(e)$ is termed the Hansen coefficient.

Equations (\ref{tf}) and (\ref{hanseneq}) imply:
\begin{equation}
 \mat S=
    \frac{3G m_0}{2a^3} \sum_{l=-2}^2\sum_{k=-\infty}^\infty
   \erm^{i\{t(k n-l\omega)+ l\,\varpi\}}
   \mat Y_lU_{kl}\label{Saap2}
\end{equation}
where $U_{k,-1}=U_{k,1}=0$ and 
\begin{equation}
  U_{k,-2}=\frac{X^{-3,-2}_k}{\sqrt 2}\,,  \ \ 
 U_{k0}=\frac{X^{-3,0}_k}{\sqrt 3}\,,  \ \ 
 U_{k2}=\frac{ X^{-3,2}_k}{\sqrt 2} \,.\label{Uk02}
\end{equation}
The symmetry property $X^{n^\prime,-m}_{-k} = X^{n^\prime,m}_{k}$ implies
\begin{equation}
  U_{kj}=U_{-k,-j}\,.\label{symUk}
\end{equation}

The centrifugal force in equation (\ref{F}) can be represented as
\begin{equation}
  \mat C = \frac{\omega^2}{\sqrt 3}\mat Y_0\,. \label{C2}
\end{equation}

To obtain the almost periodic solution of the deformation equation
\begin{equation}
  \tau\dot{\widetilde{\mathbf{B}}} + \widetilde{\mat B} = \mathbf{C}+\mathbf S,\label{defeq}
\end{equation}
solving for each Fourier mode separately suffices. An alternative approach involves using the variation of constants formula:
 \begin{equation}\begin{split}
   \widetilde{\mat B}(t)&=\mat B_d(t):=\int_{-\infty}^0\frac{\erm^{s/\tau}}{\tau}
   \frac{ \mat C +\mat S(t+s)}{\gamma+\alpha}ds= \frac{ \mat C}{\gamma+\alpha}\\ & +
     \frac{3G m_0}{2a^3} \sum_{l=-2}^2\sum_{k=-\infty}^\infty
   \erm^{i\{t(k n-l\omega)+ l\,\varpi\}}
  \frac{1}{(\gamma+\alpha)\big(1+i (k n-l\omega)\big)} \mat Y_lU_{kl}\\  &=k_\circ
   \frac{R^5 \omega^2}{G\Io} \frac{\mat Y_0}{3\sqrt 3}+
   \frac{m_0 R^5}{2\Io a^3}\sum_{l=-2}^2\sum_{k=-\infty}^\infty
  \erm^{i\{t(k n-l\omega)+ l\,\varpi\}} 
  k_2(k n-l\omega)  \mat Y_lU_{kl}\\  &=k_\circ
  \zeta_c \frac{\mat Y_0}{3\sqrt 3}+
  \zeta_{\scriptscriptstyle T} \sum_{l=-2}^2\sum_{k=-\infty}^\infty
   \erm^{i\{t(k n-l\omega)+ l\,\varpi\}}
  k_2(k n-l\omega)  \mat Y_lU_{kl}.
   \end{split}\label{DFWb}
 \end{equation}
 Here, the definitions of the Love number $k_2$ and the secular Love number $k_\circ$ from equation (\ref{lov}) are used as well as the definitions of $\zeta_c$ and $ \zeta_{\scriptscriptstyle T}$ from
 equation (\ref{zeta}).

Given that \[\int_{-\infty}^0\frac{\erm^{s/\tau}}{\tau}ds=1,\]
this formula indicates that the almost periodic solution of the tide equation is a time-averaged tidal force with an exponential weight decaying towards the past, characterized by time $\tau$. Note that when $\tau>0$ is nearly zero, integration by parts  of the right-hand side of equation \eqref{DFWb} yields
 \begin{equation}
   \mat B_d(t)-k_\circ
  \zeta_c \frac{\mat Y_0}{3\sqrt 3}
   \approx  \frac{\mat S(t)}{\gamma+\alpha}
   -\tau  \frac{\dot{\mat S}(t)}{\gamma+\alpha}\approx
   \frac{\mat S(t-\tau)}{\gamma+\alpha}. \label{lag}
 \end{equation}
This represents the usual time delay approximation with corrections of the order of $\tau^2$.

The limit case of $\tau\to\infty$ also presents interest. Here, we can interpret the averaging in equation (\ref{DFWb}) as approximately the ordinary averaging
\[\lim_{\tau\to\infty}\frac{1}{\tau}\int_{-\tau}^0 \frac{\mat S}{\gamma+\alpha} ds.\]

\section{Deformation  Manifold.}

\label{avsec2}

The function \(t\to\mat B_d\) provides a solution to the deformation equation \eqref{defeq}
only when \(\epsilon_d=0\). To analyze the case where $\epsilon_d>0$,
we introduce new deformation variables \(\boldsymbol{\delta}\mat B\):
\begin{equation}
    \widetilde{\mat B} = \mat B_d+\boldsymbol{\delta}\mat B,
\end{equation}
and using these variables we  write equation (\ref{epd})
\begin{equation}
\begin{split}  
\ddot {\mathbf x} &= G(m_0+m) \left\{
- \frac{\mathbf x}{|\mathbf x|^3}
+\epsilon_d\frac{\Io}{m}\left(-\frac{15}{2}\frac{1}{|x|^7} 
(\widetilde{\mathbf b} \mathbf x \cdot\mathbf x)\mathbf x  
+3\frac{1}{|\mathbf x|^5} \widetilde{\mathbf b}\mathbf x\right)
\right\}\\
\dot\ell_s &=
-\epsilon_d\frac{3G \Io m_0}{\|\mat x\|^5}\left\{
x_1x_2 (\tilde b_{ 22}-\tilde b_{ 11})+\tilde b_{12} (x_1^2-x_2^2)\right\}\\
\ell_s &= \omega \Io(1-\epsilon_d \tilde b_{3 3})\\
\tau\dot{\boldsymbol{\delta}\mat B} &+ \boldsymbol{\delta}\mat B = \Oc(\ep_d).
\end{split}\label{epd4}
\end{equation}

\begin{figure}[hbt!]
\centering
\includegraphics[scale=0.5]{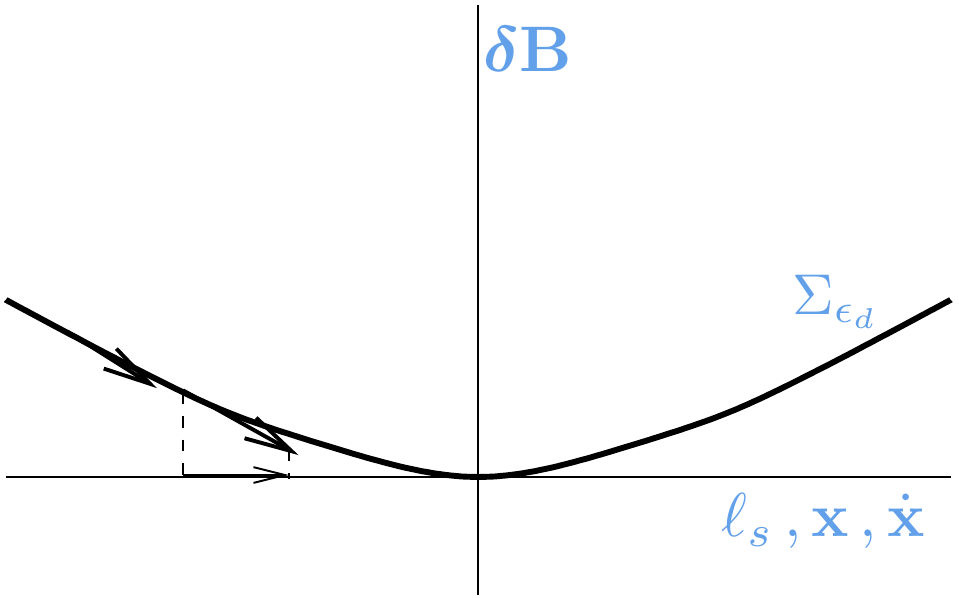}
\caption{Illustration of the Deformation Invariant Manifold \( \Sigma_{\ep_d}:=\Big\{(\mat x,\dot{\mat x},\ell_s,\ep_d)\to\boldsymbol{\delta}\mat B\Big\}\). With the parameterization defined by \((\mat x,\dot{\mat x},\ell_s,\ep_d)\), the vector field on \(\Sigma_{\ep_d}\) follows from (\ref{epd5}).}
\label{Sg}
\end{figure}

For \(\ep_d=0\), equation (\ref{epd2}) possesses the invariant manifold:
\begin{equation}
  \Sigma_0:=\{\boldsymbol{\delta}\mat B=0\}.
\end{equation}
The variables \(\boldsymbol{\delta}\mat B\) are transversal to \(\Sigma_0\), and all associated eigenvalues equal \(-1/\tau<0\). Given this, a theorem by Fenichel \cite[Theorem 3]{fenichel1971persistence} suggests that for sufficiently small \(\ep_d\), there is an invariant manifold represented as a graph:
\begin{equation}
  \Sigma_{\ep_d}:=\Big\{(\mat x,\dot{\mat x},\ell_s,\ep_d)\to\boldsymbol{\delta}\mat B\Big\}.
\end{equation}
Additionally, \(\Sigma_{\ep_d}\) approximates \(\Sigma_0\) to order \(\ep_d\), as visualized in Figure \ref{Sg}. The vector field on \(\Sigma_{\ep_d}\), considering corrections of order \(\ep_d\), is derived from equations (\ref{epd4}) by ignoring the variables \(\boldsymbol{\delta}\mat B\) and setting \( \widetilde{\mat B} = \mat B_d\) in the equations for \(\dot{\mat x}\) and \(\ell\). Thus, the equation on \(\Sigma_{\ep_d}\) is:
\begin{equation}
\begin{split}  
\ddot {\mathbf x} &= G(m_0+m) \left\{
- \frac{\mathbf x}{|\mathbf x|^3}
+\epsilon_d\frac{\Io}{m}\left(-\frac{15}{2}\frac{1}{|x|^7} 
({\mathbf b}_d \mathbf x \cdot\mathbf x)\mathbf x  
+3\frac{1}{|\mathbf x|^5} {\mathbf b}_d\mathbf x\right)
\right\}\\
\dot\ell_s &=
-\epsilon_d\frac{3G \Io m_0}{\|\mat x\|^5}\left\{
x_1x_2 ( b_{d 22}- b_{d 11})+b_{d12} (x_1^2-x_2^2)\right\}\\
\ell_s &= \omega \Io(1-\epsilon_d  b_{d 3 3}),
\end{split}\label{epd5}
\end{equation}
where, \(\mat b_d=\mat R(\phi)\mat B_d\mat R^{-1}(\phi)\).

The Fenichel theorem requires a specific condition concerning the eigenvalues of the linear equation: they must be sufficiently distant from the imaginary axis, depending on the flow on $\Sigma_0$, which is fulfilled in this case since they are constant.

 When $n$ and $\omega$ are neither small, to ensure the validity of the averaging, nor excessively large, which would violate inequalities \eqref{zeta} and result in large deformations, the approximation of $\Sigma_0$ by $\Sigma_{\ep_d}$ remains accurate. Under these conditions, changes in the Keplerian elements and spin are gradual, allowing the body ample time to adjust. The body maintains an average shape consistent with its secular Love number; for $\alpha = 0$, it remains in hydrostatic equilibrium, countering centrifugal forces and slow tides.

 An intriguing scenario arises when either $\tau n\gg 1$ or $\tau \omega\gg 1$.
 Here, the body lacks the time to relax amid orbital and spin modifications, causing the deformation to retain a memory of a past initial state. In such situations, Fenichel's theorem is not applicable. If $\tau \gg 1$ and the initial condition is $\widetilde {\mat B}=\widetilde{\mat B}_{\circ}$, the solution to the homogeneous equation $\tau\dot{\widetilde{\mat B}} +\widetilde{\mat B}= 0$ decays slowly as
\begin{equation}
 \widetilde{\mat B}(t)=\widetilde{\mat B}_{\circ}\erm^{-t/\tau}\,.
\end{equation}
In \cite{ragazzo2022librations}, in a situation similar to this one, we added a permanent deformation $\widetilde{\mat B}_{\circ}$ to $\mat B_d$ and continued. Adopting the same approach here is feasible, even without a mathematical basis. However, we must separate the orbital motion's averaging into two components: one for terms with $\mat B_d$ and another for terms with $\widetilde{\mat B}_{\circ}$. The averaging of terms associated with $\widetilde{\mat B}_{\circ}$ would resemble the averaging in rigid body problems.
Here, we will not introduce the permanent deformation to keep the following analysis as simple as
possible.

Later in this paper, we'll explore situations where $\tau n$ is large, assuming that, despite its size, Fenichel's conditions remain met. This assumption warrants further mathematical scrutiny, potentially through multi-timescale system theories.

\section{Orbital Averaging}
\label{avesec3}

We average equation (\ref{epd5}) with respect to orbital motion. We set the scaling parameter $\ep_d$ to 1. Equations (\ref{epd5}) and (\ref{DFWb}) then become:
\begin{equation}
\begin{split}  
\ddot {\mathbf x} &= G(m_0+m) \left\{
- \frac{\mathbf x}{|\mathbf x|^3}
+\frac{\Io}{m}\left(-\frac{15}{2}\frac{1}{|x|^7} 
({\mathbf b}_d \mathbf x \cdot\mathbf x)\mathbf x  
+3\frac{1}{|\mathbf x|^5} {\mathbf b}_d\mathbf x\right)
\right\},\\
\dot\ell_s &=
-\frac{3G \Io m_0}{\|\mat x\|^5}\left\{
x_1x_2 ( b_{d 22}- b_{d 11})+b_{d12} (x_1^2-x_2^2)\right\},\\
\ell_s &= \omega \Io(1-  b_{d 3 3}),\\
\mat b_d & =k_\circ
  \zeta_c \frac{\mat Y_0}{3\sqrt 3}+
  \zeta_{\scriptscriptstyle T} \sum_{l=-2}^2\sum_{k=-\infty}^\infty
 \erm^{i\{tk n+ l\,\varpi\}}
  k_2(k n-l\omega)  \mat Y_lU_{kl}.
\end{split}\label{epd6}
\end{equation}

Using variables $\ell$, $\mat A$, and $f$ defined in equations (\ref{laplace}) and (\ref{reduced}), equation (\ref{epd6}) transforms to: 
\begin{equation}\begin{split}
\dot{\ell}_s &=
- 3c  \frac{ \Io}{m}   E_{1} ,\\
\dot{\ell} &=
3 c  \frac{ \Io}{m}  E_{1},\\
\ell_s &= \omega \Io(1- \langle b_{d33}\rangle),\\
  \dot {\mat A}&= 3  \, \frac{\ell}{\mu}\, \frac{\Io}{m}  \bigg(-\frac{5}{2}\mat E_{2}   
  +\mat E_{3}\bigg)\times\mat e_3  +3   \frac{c }{\ell}\frac{\Io}{m}\mat E_{4}+
  3 \, \frac{c}{\ell}\frac{ \Io}{m} E_{1} \mat A.
   \end{split}\label{eq2}
\end{equation}

The terms requiring averaging are:
\begin{equation}
  \begin{split}
    E_{1}&=   \left\langle\frac{
x_1x_2 ( b_{d 22}- b_{d 11})+b_{d12} (x_1^2-x_2^2)}{\|\mat x\|^5}\right\rangle, \\
 \mat E_{2}&=   \left\langle \frac{1}{|\mat x|^7}\left(
   \mathbf b_b  \mathbf x \cdot\mathbf x\right)\mathbf x\right\rangle,\\
 \mat E_{3}&= \left\langle\frac{1}{|\mathbf x|^5}  \mathbf b_d \mathbf x\right\rangle,\\
 \mat E_{4}&= \left\langle \frac{1}{|\mat x|^5}
   \Big(\big(\mathbf{x}\times\mat b_d\,\mathbf{x}\big)
   \cdot \mat e_3\Big)\frac{\mat x}{|\mat x|}\right\rangle,\\
 \langle b_{d33}\rangle&= \left\langle k_\circ
  \zeta_c \frac{\mat e_3\cdot\mat Y_0\mat e_3}{3\sqrt 3}+
  \zeta_{\scriptscriptstyle T} \sum_{k=-\infty}^\infty
\erm^{itk n}
k_2(k n) U_{k0} (\mat e_3\cdot\mat Y_0\mat e_3)\right\rangle,
\end{split}\label{Ea}
\end{equation}
where $\langle h\rangle=\frac{1}{2\pi}\int_0^{2\pi}h(M)dM$ represents the average over the mean anomaly.

The total angular momentum is conserved and given by:
  \begin{equation}
    \ell_{\scriptscriptstyle T}:=\ell+\ell_s. \label{lcons}
  \end{equation}

The averaged result yields:

\noindent \textit{The term $E_{1}$:}
\[
\begin{split}
E_{1} &= \left\langle\frac{x_1x_2(b_{d 22}- b_{d 11}) + b_{d12}(x_1^2-x_2^2)}{\|\mat x\|^5}\right\rangle \\
&= \sum\limits_{k=-\infty}^\infty \frac{i \zeta_{\scriptscriptstyle T} \left(X^{-3,-2}_k X^{-3,2}_{-k}  k_2(k n+2\omega) - X^{-3,-2}_{-k} X^{-3,2}_{k} k_2(k n-2\omega)\right)}{2 a^3} \\
&= \frac{i \zeta_{\scriptscriptstyle T}}{2 a^3} \sum\limits_{k=-\infty}^\infty X^{-3,2}_{k} X^{-3,2}_{k} \left(k_2(-k n+2\omega) - k_2(k n-2\omega)\right) \\
&= \frac{\zeta_{\scriptscriptstyle T}}{a^3} \sum\limits_{k=-\infty}^\infty \Big(X^{-3,2}_{k}(e)\Big)^2 \mathrm{Im} \, k_2(k n-2\omega)
\end{split}
\]
where we used,
from equation (\ref{lov}), that $k_2(-\sigma)$ is the complex conjugate of $k_2(\sigma)$,
represented as $\overline{k_2(\sigma)}$.

We write $E_{1}$ as
\begin{equation}
    E_{1} = \frac{\zeta_{\scriptscriptstyle T}}{a^3} \mathcal{A}_0\,,\quad \mathcal{A}_0 = \sum\limits_{k=-\infty}^\infty \Big(X^{-3,2}_{k}(e)\Big)^2 \mathrm{Im} \, k_2(k n-2\omega).
\label{E1a}
\end{equation}

\noindent \textit{The terms $\left(-\frac{5}{2}\mat E_{2} + \mat E_3\right)$:} 
The calculation of these terms resembles that of $E_1$. The analysis was extended and performed using the software ``Mathematica''. We will skip the detailed steps. The outcomes are:
\begin{equation}
    \mathcal{A}_{1} = -\sum_k (X_{k}^{-4,1} + 5 X_{k}^{-4,3}) X_{k}^{-3,2} \mathrm{Re}\, k_2(n k-2\omega) + 2X_{k}^{-4,1} X_{k}^{-3,0} \mathrm{Re}\, k_2(nk)
\end{equation}    

\begin{equation}
    \mathcal{A}_{2} = \sum_k (5 X_{k}^{-4,3} - X_{k}^{-4,1}) X_{k}^{-3,2} \mathrm{Im}\, k_2(kn -2\omega) + 2 X_{k}^{-4,1} X_{k}^{-3,0} \mathrm{Im}\, k_2(nk)
\end{equation} 

\begin{equation}
   \mathcal{A}_{3} = X_{0}^{-4,1}
\end{equation}
and
\begin{equation}
\begin{split}
\left(\begin{matrix} \Big(-\frac{5}{2}\mat E_{2} + \mat E_{3}\Big)_1 \\
\Big(-\frac{5}{2}\mat E_{2} + \mat E_{ 3}\Big)_2 \end{matrix} \right) 
&= \left\{\frac{\zeta_{\scriptscriptstyle T}}{4 a^4} \left(\begin{matrix} \mathcal{A}_1 & -\mathcal{A}_2 \\ \mathcal{A}_2 & \mathcal{A}_1 \end{matrix} \right) - \frac{k_\circ \zeta_c}{6 a^4} \mathcal{A}_3 \right\} \left(\begin{matrix} \cos\varpi\\ \sin\varpi \end{matrix} \right) \\
&= \left\{\frac{\zeta_{\scriptscriptstyle T}}{4 a^4} \mathcal{A}_1 - \frac{k_\circ \zeta_c}{6 a^4} \mathcal{A}_3 \right\} \mat e_A + \frac{\zeta_{\scriptscriptstyle T}}{4 a^4} \mathcal{A}_2 \mat e_H\,,
\end{split}
\end{equation}
where we used equations (\ref{orbp}).

\noindent \textit{The term $\mat E_{4}$:} 
Detailed steps are omitted as before. The outcomes are:

\begin{equation}
    \mathcal{A}_{4} = \sum_k X_{k}^{-3,2} (X_{k}^{-3,1} + X_{k}^{-3,3}) \mathrm{Im} \, k_2(kn -2\omega)
\end{equation}    

\begin{equation}
    \mathcal{A}_{5} = \sum_k (X_{k}^{-3,3} - X_{k}^{-3,1}) X_{k}^{-3,2} \mathrm{Re} \, k_2(kn-2\omega)
\end{equation} 
and 
\begin{equation}
  \mat E_{4}= \left(\begin{matrix} \mat E_{4} \cdot \mat e_1 \\ \mat E_{4} \cdot \mat e_2 \end{matrix} \right) = \frac{\zeta_{\scriptscriptstyle T}}{2 a^3} \left( \mathcal{A}_4 \mat e_A + \mathcal{A}_5 \mat e_H \right)\,.
\end{equation}

\nd {\it The term $   
  \langle b_{d33}\rangle$:}
\begin{equation}\begin{split}
   \langle b_{d33}\rangle&= \left\langle k_\circ
  \zeta_c \frac{\mat e_3\cdot\mat Y_0\mat e_3}{3\sqrt 3}+
  \zeta_{\scriptscriptstyle T} \sum_{k=-\infty}^\infty
\erm^{itk n}
k_2(k n) U_{k0} (\mat e_3\cdot\mat Y_0\mat e_3)\right\rangle\\
&= -\frac{2}{3}k_\circ\left(\frac{\zeta_c}{3}+ \frac{\zeta_{\scriptscriptstyle T}}{(1-e^2)^{3/2}}
  \right)
\end{split}
\end{equation}
where we used that $X^{-3,0}_0=(1-e^2)^{-3/2}$ \cite{laskar2010explicit}\footnote{The gravity field  coefficient $J_2$ (dynamic form factor) is related to  $\Io \langle b_{d33}\rangle$
by means of $\Io \langle b_{d33}\rangle=-\frac{2}{3}mR^2 J_2$ that implies
\begin{equation}
J_2=\frac{\Io}{m R^2}k_\circ\left(\frac{\zeta_c}{3}+ \frac{\zeta_{\scriptscriptstyle T}}{(1-e^2)^{3/2}}
  \right)
\end{equation}}.

For the Kepler problem, the following relations hold: 
\begin{equation}
    \ell^2=\mu ca(1-e^2)\Rightarrow
    (1-e^2)=\frac{\ell^2}{\mu ca}\,.\label{lmca}
\end{equation}
Assuming $\ell>0$, we can use  \( G(m_0+m) = n^2a^3 \)
 to write:
\begin{equation}
    \frac{\ell}{\mu a^2} = n \sqrt{1-e^2}\,.\label{lmca3}
\end{equation}
  
Using the above relations, further calculations yield:
\begin{equation}
\begin{split}
    \dot{\mat A}
    & = \frac{3c}{2\ell} \frac{ \Io}{m} \frac{\zeta_{\scriptscriptstyle T}} {a^3}
    \bigg\{ \frac{1-e^2}{2}  {\cal A}_2
    + {\cal A}_4
    + 2e\,{\cal A}_0 \bigg\}\, \mat e_A \\
    & + \frac{3c}{2\ell} \frac{ \Io}{m} \frac{\zeta_{\scriptscriptstyle T}} {a^3}
    \bigg\{ {\cal A}_5 - \frac{1-e^2}{2}  {\cal A}_1
    \bigg\}\mat e_H + \frac{\Io}{m}\frac{k_\circ\zeta_c }{6 a^4}
    {\cal A}_3\mat e_H.
\end{split}
\end{equation}

Given that \( \mat A = e\, \big(\cos\varpi \mat e_1+\sin\varpi \mat e_2\big) = e \,\mat e_A \) and \( \dot {\mat e}_A = \dot \varpi \mat e_H \), we deduce:
\begin{equation}
    \dot{\mat A} = \dot e\,\mat e_A + \dot \varpi\,e\, \mat e_H \,.
\end{equation}
Thus, the final averaged equations are:

\begin{equation}
   \begin{split}
   \dot e&=\frac{3c}{2\ell} \frac{ \Io}{m} \frac{\zeta_{\scriptscriptstyle T}} {a^3}
   \bigg\{ \frac{ 1-e^2}{2}  {\cal A}_2
+{\cal A}_4+
2e\,{\cal A}_0 \bigg\}\\
e\dot\varpi&=
 \frac{3c}{2\ell} \frac{ \Io}{m} \frac{\zeta_{\scriptscriptstyle T}} {a^3}
   \bigg\{ {\cal A}_5-\frac{ 1-e^2}{2}  {\cal A}_1
 \bigg\}+\frac{ \Io}{m}\frac{k_\circ\zeta_c }{6 a^4}
 {\cal A}_3\\
 \dot \ell&=  3c  \frac{ \Io}{m}\frac{\zeta_{\scriptscriptstyle T}} {a^3}{\cal A}_0\\
  \dot{\ell}_s & = - 3c  \frac{ \Io}{m}\frac{\zeta_{\scriptscriptstyle T}} {a^3}{\cal A}_0\\
  \ell_{\scriptscriptstyle T}&=\ell+\ell_s=\text{constant}\\
 \ell_s &= \omega \Io(1-  \langle b_{d 3 3}\rangle)\\
  {\cal A}_0&=  \sum\limits_{k=-\infty}^\infty
    \Big(X^{-3,2}_{k}\Big)^2{\rm Im} \, k_2(k n-2\omega)\\
      {\cal A}_{2}&=  \sum_k
      (5 X_{k}^{-4,3}-X_{k}^{-4,1}) X_{k}^{-3,2}{\rm Im}\, k_2(kn -2\omega)+
      2 X_{k}^{-4,1} X_{k}^{-3,0}{\rm Im}\, k_2(nk)\\
   {\cal A}_{4}&=  \sum_kX_{k}^{-3,2} (X_{k}^{-3,1}+X_{k}^{-3,3})
   {\rm Im }\,k_2(kn -2\omega)\\
   {\cal A}_{1}&=-
    \sum_k(X_{k}^{-4,1}+5 X_{k}^{-4,3}) X_{k}^{-3,2}
     {\rm Re}\, k_2(n k-2\omega)+2X_{k}^{-4,1} X_{k}^{-3,0}
     {\rm Re}\, k_2(nk)\\  {\cal A}_{3}&=
     X_{0}^{-4,1}\\ {\cal A}_{5}&= \sum_k(X_{k}^{-3,3}-X_{k}^{-3,1}) X_{k}^{-3,2}
     {\rm Re}\, k_2(kn-2\omega)\\
      \langle b_{d33}\rangle
&= -\frac{2}{3}k_\circ\left(\frac{\zeta_c}{3}+ \frac{\zeta_{\scriptscriptstyle T}}{(1-e^2)^{3/2}}
\right)\\
  \zeta_c &= \frac{R^5 \omega^2}{G\Io}\\
  \zeta_{\scriptscriptstyle T} &= \frac{m_0 R^5}{2\Io a^3}\\
  \mu&=\frac{m_0 m}{m_0+m}\\
  c&=G m m_0\\
  n^2a^3&= G(m_0+m) \\
    \frac{\ell}{\mu a^2} &= n \sqrt{1-e^2}\,.
\end{split}
\label{av}
\end{equation}

\subsection{Computation of Hansen coefficients}
\label{hansensec}

The Hansen coefficients depend solely on the eccentricity. Following \cite{cherniack1972computation}, we express, for \(n<0\) and \(m \ge 0\),
\begin{equation}
  \left(\frac{r}{a}\right)^n\erm^{i m f}=
  \left(\sum_{k=-\infty}^\infty X^{-1,0}_k(e) \erm^{i k M}\right)^{|n|}\,
\left( \sum_{l=-\infty}^\infty X^{0,1}_l(e) \erm^{i l M}\right)^m\,.
 \label{hanseneq2}
\end{equation}
Thus, to compute any series \(\left(\frac{r}{a}\right)^n\erm^{i m f}\), one can employ series multiplication of the fundamental series of \(a/r\) and \(\erm^{if}\). This multiplication can be efficiently executed with an algebraic manipulator.

For the computation of the series for \(\frac{a}{r}\) and \(\erm^{if}\), one can refer to
\cite{Murray} Section 2.5:
\begin{equation}
\begin{split}
\frac{a}{r}&=\frac{e \cos f+1}{1-e^2}\\
\cos f&=\frac{\erm^{if}+\erm^{-if}}{2}=
-e+ 2\frac{1-e^2}{e}\sum\limits_{k=1}^\infty J_k(ke)\cos(k M)\\
\sin f&=\frac{\erm^{if}-\erm^{-if}}{2i}=2\sqrt{1-e^2}\sum\limits_{k=1}^\infty \frac{1}{k}\frac{d}{de}J_k(ke)\sin(k M)\\
J_k(x)&=\frac{1}{k!}\left(\frac{x}{2}\right)^k\sum\limits_{l=0}^\infty(-1)^l
\frac{\left(\frac{x}{2}\right)^{2l}}{l!(k+1)(k+2)\ldots(k+l)}\,,
\end{split}
\end{equation}
where \(J_k(x)\) denotes the Bessel function. The series for \(J_k(x)\) converges absolutely for all values of \(x\).

Up to second order in eccentricity and with $\erm^{iM}=z$ the fundamental series are:
\begin{equation}
  \begin{split}
    \frac{r}{a}&= 1- e\frac{1}{2} \left(z+z^{-1}\right)+ e^2 \left(\frac{1}{2}-\frac{1}{4} \left(z^2+z^{-2}\right)\right)+\Oc(e^3)\\
    \frac{a}{r}&=1+e\frac{1}{2} \left(z+z^{-1}\right)+e^2\frac{ \left(z^2+z^{-2}\right)}{2}+\Oc(e^3)\\
     \erm^{if}&=z\left\{1+e \left(z-z^{-1}\right)+e^2 \left(\frac{9 z^2}{8}-1-\frac{z^{-2}}{8}\right)\right\}+\Oc(e^3)\\
       \erm^{-if}&= z^{-1}\left\{1+e \left(z^{-1}-z\right)+e^2 \left(\frac{9z^{-2}}{8 }-1-\frac{z^2}{8}\right)
       \right\}+\Oc(e^3)\\\\
\end{split} \end{equation}
These expressions and equation \eqref{hanseneq2} imply $X^{n,m}_k=\Oc (e^{|m-k|})$.

\subsection{The equations in \cite{correia2022tidal}}

Some relations between the Hansen coefficients presented in \cite{correia2022tidal}, equations (158) and (159), are:
\begin{equation}
\begin{split}
  \sqrt{1-e^2} kX^{-3,0}_k &= \frac{3}{2}e (X^{-4,1}_k-X^{-4,-1}_k) = \frac{3}{2}e (X^{-4,1}_k-X^{-4,1}_{-k}),\\
  \sqrt{1-e^2} kX^{-3,2}_k &= \frac{e}{2} (5X^{-4,3}_k-X^{-4,1}_k) + 2X^{-4,2}_k,\\
  X^{-3,3}_k &= \frac{1}{e} \left( 2(1-e^2)X^{-4,2}_k-2X_k^{-3,2} -e X^{-3,1}_k \right) \,.
\end{split}
\end{equation}

One can use these equations to simplify equations (\ref{av}). After such simplifications, the equation governing the eccentricity is:

\begin{equation}
\begin{split}
 \dot e &= \frac{3c}{2\ell} \frac{\Io}{m} \frac{\zeta_{\scriptscriptstyle T}} {a^3} \frac{ 1-e^2}{3e}
\sum_{k=-\infty}^{\infty} \left\{ k \sqrt{1-e^2} (X^{-3,0}_k)^2 {\rm Im} \, k_{2}(nk) \right. \\
 & \left. -3 \left(2-k\sqrt{1-e^2}\right) (X^{-3,2}_{k})^2 {\rm Im} \, k_2(k n-2\omega) \right\} \,.
\end{split}
\label{secLuc9}
\end{equation}
For further simplification, one can apply \(\ell = \mu \sqrt{G(m+m_0)a(1-e^2)}\), yielding:

\begin{equation}
\begin{split}
 \dot e &= n \frac{m_{0}}{m} \frac{R^5}{a^5} \frac{\sqrt{1-e^2}}{4e}
\sum_{k=-\infty}^{\infty} \left\{ k \sqrt{1-e^2} (X^{-3,0}_k)^2 {\rm Im}\, k_{2}(nk) \right. \\
 & \left. -3 \left(2-k\sqrt{1-e^2}\right) (X^{-3,2}_{k})^2 {\rm Im} \, k_2(k n-2\omega) \right\} \,.
\end{split}
\end{equation}
This result corresponds to equation (129) in \cite{correia2022tidal}.

Our expression for the variation of the longitude of the periapsis, \(\dot \varpi\), differs from equation (130) in \cite{correia2022tidal} due to the neglect of centrifugal deformation in the cited work.

\section{Averaged Equations: A Geometrical Approach}
\label{singular}

In the following two sections, we analyze equation (\ref{av}) from a geometric perspective using
singular perturbation theory.

The longitude of the periapsis, $\varpi$, is absent from the equation  for $ \dot e $ in \eqref{av}.
Therefore, the dynamics of the state variables $e, \ell,$ and $\ell_s$ can be analyzed independently of $\varpi$.
The conservation of total angular momentum, $ \ell_{\scriptscriptstyle T} = \ell + \ell_s$,
implies that it is sufficient to observe the dynamics of $e$ and $\ell_s$.

While the dynamics unfolds within  two-dimensional surfaces, on the level sets of angular momentum, analyzing the equations within a three-dimensional phase space proves more insightful. This approach facilitates a comprehensive understanding of the global dynamics and the impact of varying angular momentum. After some investigation, we selected $(\omega, e, a)$ as the phase-space variables, with $n = \sqrt{\frac{G(m + m_0)}{a^3}}$ being a derived quantity. The differential equation for $a=\frac{\ell^2}{\mu c (1-e^2)}$ is obtained from the equations for
$\dot\ell$ and $\dot e$.
Henceforth we use the approximation
\[ \ell_s = \omega \Io(1 - \langle b_{d 3 3}\rangle) \approx \omega\Io\,.\]

Equations \eqref{av} and the identity $\frac{ 3c}{mn^2a^3}=\frac{3 m_0}{m+m_0}$ imply
\begin{equation}
  \begin{split}
   \frac{\dot \omega}{n^2}& = -\left(\frac{3 m_0}{m+m_0}\zeta_{\scriptscriptstyle T}\right) {\cal A}_0\\ 
   \frac{\dot e}{n}&= \left(\frac{3 m_0}{m+m_0}\zeta_{\scriptscriptstyle T}\right)
 \frac{\Io }{\mu a^2} \frac{1}{2 \sqrt{1-e^2}} 
   \bigg\{ \frac{ 1-e^2}{2}  {\cal A}_2
+{\cal A}_4+
2e\,{\cal A}_0 \bigg\}\\
\frac{\dot{a}}{n} &
= \left(\frac{3 m_0}{m+m_0}\zeta_{\scriptscriptstyle T}\right)\frac{\Io }{\mu a^2}
\frac{a}{(1-e^2)^{3/2}}\left\{ e\left(\frac{ 1-e^2}{2}  {\cal A}_2 + {\cal A}_4 \right)+
2\,{\cal A}_0 \right\}\, .
 \end{split}
\label{av2}
\end{equation}

Conservation of angular momentum 
$\ell_{\scriptscriptstyle T} =   \sqrt{\mu c a} \sqrt{1-e^2}  + \Io \omega$ implies
\begin{equation}
  \frac{\omega}{n}= \frac{\ell_{\scriptscriptstyle T}}{\Io n}
  \left(1 -\sqrt{\frac{a \mu c}{\ell^2_{\scriptscriptstyle T}}}\sqrt{1-e^2}\right).
  \label{lt1}
\end{equation}
This suggests   the following nondimensionalization of $a$: 
\begin{equation}
  \tilde{a} := \frac{a}{a_\circ}\,, \quad \text{where} \quad
  a_\circ := \frac{\ell_{\scriptscriptstyle T}^2}{\mu c}
\label{secLuc5}
\end{equation}
is defined as the radius of the circular orbit for two point masses, $m_0$ and $m$, possessing an orbital angular momentum of $\ell = \ell_T$.

Let
\begin{equation}
  n_\circ=\frac{\ell_{\scriptscriptstyle T}}{\mu a_\circ^2}=\frac{c^2 \mu}{\ell_{\scriptscriptstyle T}^3}\label{ncirc}
\end{equation}
be the angular frequency of  the circular orbit of  radius $a_\circ$. Kepler's third law implies, $n^2a^3=G(m+m_0)=n_\circ^2a_\circ^3$ and
so 
\begin{equation}
  n=n_\circ\,\frac{1}{\tilde a^{3/2}}\,.\label{ncirc2}
\end{equation}

Conservation of angular momentum, as expressed in equation (\ref{lt1}),
implies
\begin{equation}
 \frac{\omega}{n} = \epsilon^{-1} \tilde{a}^{\frac{3}{2}}(1-\tilde{a}^{\frac{1}{2}} \sqrt{1-e^2})\,,
\label{secLuc13}
\end{equation}
where
\begin{equation}
\epsilon := \frac{\Io}{\mu a_\circ^2} = \frac{\Io n_\circ}{\ell_{\scriptscriptstyle T}}=  \frac{\Io \mu c^2}{\ell_{\scriptscriptstyle T}^4}\,.
\label{secLuc7}
\end{equation}
For the Mercury-Sun system, where $m_0$ is the mass of the Sun, $\epsilon = 6.8 \times 10^{-10}$, and for the Earth-Moon system, where $m_0$ is the mass of the Moon, $\epsilon = 0.0036$. Although $\epsilon$ appears to be very small for all problems of interest, in this section, we will conduct a geometric analysis with an arbitrary value of $\epsilon$ to elucidate the global properties of the equations.

Using the above definitions  equations (\ref{av2}) can be written in nondimensional form as
\begin{equation}
  \begin{split}
   \frac{\dot \omega}{n_\circ^2}& = -N\frac{1}{\tilde a^6} {\cal A}_0\\ 
   \frac{\dot e}{n_\circ}&= \epsilon\,N 
  \frac{1}{2 \tilde a^{13/2}\sqrt{1-e^2}} 
   \bigg\{ \frac{ 1-e^2}{2}  {\cal A}_2
+{\cal A}_4+
2e\,{\cal A}_0 \bigg\}\\
\frac{\dot{\tilde a}}{n_\circ} &
= \epsilon\, N 
\frac{1}{\tilde a^{11/2} (1-e^2)^{3/2}}\left\{ e\left(\frac{ 1-e^2}{2}  {\cal A}_2 + {\cal A}_4 \right)+
  2\,{\cal A}_0 \right\}\\
N&= \frac{3 m_0}{m+m_0}\zeta_{\scriptscriptstyle T\circ}\quad\text{where}\quad  \zeta_{\scriptscriptstyle T\circ} = \frac{m_0 R^5}{2\Io a_\circ^3} \\
\tilde{a}&= \frac{a}{a_\circ}\,\quad \text{where}\quad
 a_\circ= \frac{\ell_{\scriptscriptstyle T}^2}{\mu c}\\
 n&=n_\circ\,\frac{1}{\tilde a^{3/2}}\quad \text{where}\quad
 n_\circ=\frac{\ell_{\scriptscriptstyle T}}{\mu a_\circ^2} \\
 \epsilon &= \frac{\Io}{\mu a_0^2} =
 \frac{\Io n_\circ}{\ell_{\scriptscriptstyle T}}=  \frac{\Io \mu c^2}{\ell_{\scriptscriptstyle T}^4}\,,
\end{split}
\label{av3}
\end{equation}
where $\mu$, $c$, $ {\cal A}_0$,  ${\cal A}_2$, and  ${\cal A}_4$ are given in equation \eqref{av}.

\subsection{Estimate of the Rate of  Spin Variations.}
\label{sect_fast}

In a time scale where the unit of time corresponds to one radian
of orbital motion, the spin angular velocity is $\omega/n$, and, from equation \eqref{av2},  the
rate of change of spin is
\[
  \frac{\dot \omega}{n^2} = -\left(\frac{3 m_0}{m+m_0}\zeta_{\scriptscriptstyle T}\right) {\cal A}_0\,,
\] 
where
\[
{\cal A}_0
  =  \sum\limits_{k=-\infty}^\infty
  \Big(X^{-3,2}_{k}(e)\Big)^2 {\rm Im} k_2(k n-2\omega)\,.
\] 

From  equation (\ref{lov})
\begin{equation}
    {\rm Im} k_2(k n-2\omega) = -k_\circ  \frac{\, \tau (k n-2\omega)}{1+ \tau^2 (k n-2\omega)^2}\,,
\label{secLuc11}
\end{equation}
and we can express
 \begin{equation}
 \begin{split}
  \frac{\dot\omega}{n^2} &= V\big(\tau n, \frac{\omega}{n}, e\big)\\
  &:= \frac{3 m_0}{(m+m_0)}\,\zeta_{\scriptscriptstyle T}\,
  k_\circ \sum\limits_{k=-\infty}^\infty
  \Big(X^{-3,2}_{k}(e)\Big)^2  \frac{\, \tau n (k -2\frac{\omega}{n})}{1+ \tau^2 n^2
    (k -2\frac{\omega}{n})^2}\,.
  \end{split}
  \label{V}
\end{equation}

\begin{figure}[hbt!]
\centering
\includegraphics[scale=0.5]{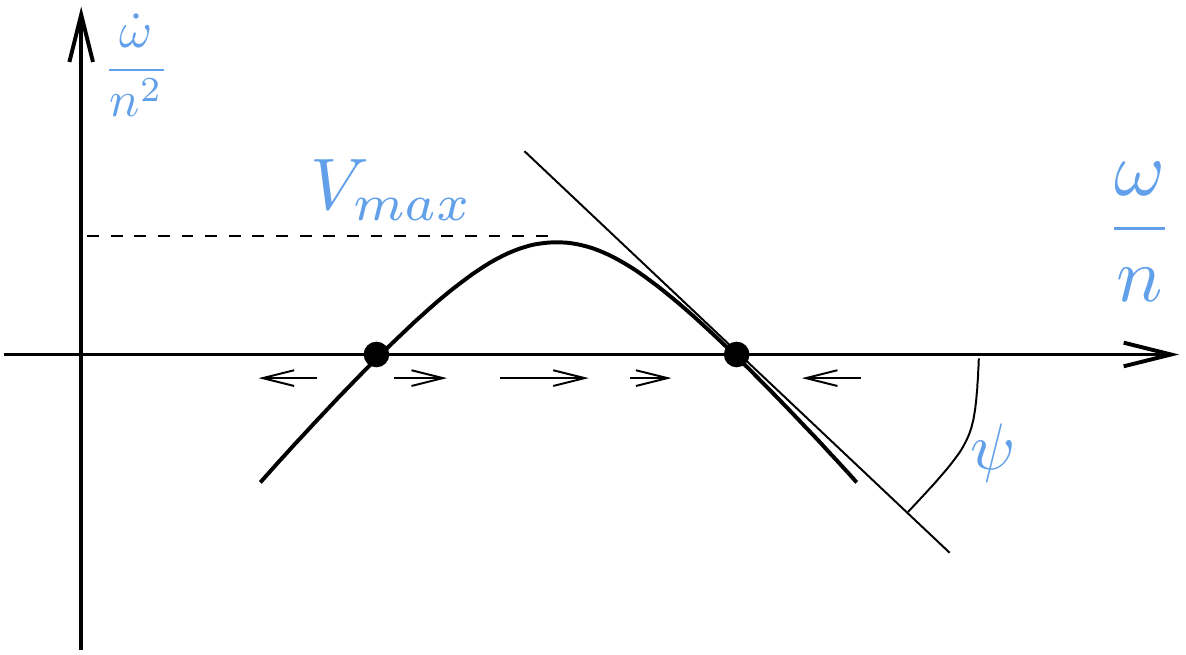}
\caption{Vector field $\frac{\dot \omega}{n^2}=V\big(\tau n, \frac{\omega}{n},e\big)$ with constant $n$ and $e$. $V_{max}$ represents the maximum rate of variation of $\frac{\omega}{n}$ and
  $\tau_s^{-1}=\tan \psi$ denotes the time constant of a stable equilibrium.}
\label{Vf}
\end{figure}

For a fixed pair $(e,n)$, $ \frac{\dot\omega}{n^2} = V\big(\tau n, \frac{\omega}{n}, e\big)$
defines a differential equation for $\frac{\omega}{n}$. We aim to estimate two typical quantities
associated with $V$:
its maximum and the time constant
near a stable equilibrium, as depicted in Figure \ref{Vf}.

The maximum value of the function $\sg\to \frac{|\sg|}{1+\sg^2}$ is $\frac{1}{2}$.
Hence,
\begin{equation}
\sum\limits_{k=-\infty}^\infty
\Big(X^{-3,2}_{k}(e)\Big)^2  \frac{\, \tau (k n-2\omega)}{1+ \tau^2 (k n-2\omega)^2}
\le \frac{1}{2}\sum\limits_{k=-\infty}^\infty
\Big(X^{-3,2}_{k}(e)\Big)^2.
\end{equation}
Applying Parseval's identity, we get
\begin{equation}
 \sum\limits_{k=-\infty}^\infty
\Big(X^{-3,2}_{k}(e)\Big)^2= \frac{1}{2\pi}\int_0^{2\pi}\frac{\erm^{i2f}}{r^3}\frac{\erm^{-i2f}}{r^3}dM=
\frac{1}{2\pi}\int_0^{2\pi}\frac{1}{r^6}dM=X^{-6,0}_0.
\end{equation}
Based on \cite{laskar2010explicit}, $X^{-6,0}_0=\frac{\frac{3 e^4}{8}+3 e^2+1}{\left(1-e^2\right)^{9/2}}$
leading to
\begin{equation}
 \sum\limits_{k=-\infty}^\infty
\Big(X^{-3,2}_{k}(e)\Big)^2  \frac{\, \tau (k n-2\omega)}{1+ \tau^2 (k n-2\omega)^2}
\le \frac{1}{2}\frac{\frac{3 e^4}{8}+3 e^2+1}{\left(1-e^2\right)^{9/2}}.\label{inhan}
\end{equation}
The right side of this inequality increases with $e$, with values:
$1/2$ for $e=0$, approximately $1.6$ for $e=0.4$, approximately $3.3$ for $e=0.5$, and approximately
$8$ for $e=0.6$. Since $\frac{m_0}{m+m_0}\le 1$,  we deduce
\begin{equation}
  \frac{\dot\omega}{n^2} \le  10 \,\zeta_{\scriptscriptstyle T}\, k_\circ \quad \text{when}\quad
  e<0.5.\label{Vmax}
 \end{equation}  
It is worth noting that $\zeta_{\scriptscriptstyle T}$, defined in equation (\ref{zeta}), is a small quantity.

For sufficiently large values of $\tau n$, the stable equilibria of $\frac{\omega}{n}$ are close to semi-integers $\frac{k}{2}$, with \(k=1,2,\ldots\),
and for these values, the dominant term in the sum of $V$ is the $k^{th}$-term \cite{correia2014deformation}. Thus, 
equation (\ref{V}) yields the time constant 
\begin{equation}
  \tau_k^{-1}\approx
  \frac{3 m_0}{(m+m_0)}\,\zeta_{\scriptscriptstyle T}\,
  k_\circ
  \Big(X^{-3,2}_{k}(e)\Big)^2  \tau n\,, 
\end{equation}
for an equilibrium  $\frac{\omega}{n}\approx \frac{k}{2}$.

Note that $V_{max}$ is independent of the characteristic time of the rheology $\tau$, whereas the time
constant $\tau_k$ has a linear dependency. A maximum rate speed $V_{max}$ proportional to
$\zeta_{\scriptscriptstyle T}\, k_\circ$
will be observed during
spin jumps. The prefactor $10$ in equation (\ref{Vmax}) varies with the eccentricity $e$. 

\subsection{Equilibria, Linearization and the
  Invariant Subspace of Zero Eccentricity.}

Using the expresions for the Hansen coefficients in Section \ref{hansensec} we can compute
the expansion of the right-hand side of equation \eqref{av3} up to first order in eccentricity:

\begin{equation}
  \begin{split}
    \frac{\dot \omega}{n_\circ^2}& = k_\circ \frac{N}{\tilde a^6}\ 
    \frac{\tau  (2 n-2 \omega )}{\tau ^2 (2 n-2 \omega )^2+1}\\ 
\frac{\dot{\tilde a}}{n_\circ} &
= - k_\circ\frac{\epsilon\, N }{\tilde a^{11/2}}\ \frac{2 \tau  (2 n-2 \omega )}{\tau ^2 (2 n-2 \omega )^2+1}\\
  \frac{\dot e}{n_\circ}&= - k_\circ\frac{\epsilon\,N} {2 \tilde a^{13/2}}\
  \frac{1}{4} e \tau n \left(\frac{6 }{n^2 \tau ^2+1}+\frac{8 \frac{\omega}{n} -8 }{4 \tau ^2 n^2
      (1-\frac{\omega}{n} )^2+1}
  \right.\\ &\left. \qquad\qquad\qquad+\frac{2 \frac{\omega}{n} -1}{\tau ^2n^2 (1-2 \frac{\omega}{n} )^2+1}
    +\frac{49 (3 -2  \frac{\omega}{n} )}{\tau ^2 n^2(3 -2 \frac{\omega}{n} )^2+1}\right)\\
    n&=n_\circ\,\frac{1}{\tilde a^{3/2}}\\
\end{split}
\label{av4}
\end{equation}

These equations imply that the plane $e=0$ is invariant.

The only equilibria of equations \eqref{av3} are on the plane $e=0$,
as shown in the next paragraph,
and are given by the curve
\begin{equation}
  \frac{\omega}{n}=1 .
\end{equation}

The equilibria of \eqref{av3} satisfiy ${\cal A}_{0} = 0$
and $\frac{ 1-e^2}{2}  {\cal A}_2 + {\cal A}_4=0$. Equation \eqref{secLuc9} shows that these equations imply
\begin{multline}
 n\sum_{k=-\infty}^{\infty} \left\{ k  (X^{-3,0}_k)^2 {\rm Im} \, k_{2}(nk)  +3 k (X^{-3,2}_{k})^2 {\rm Im} \, k_2(k n-2\omega) \right\}  \\ = \sum_{k=-\infty}^{\infty} \left\{   (X^{-3,0}_k)^2 (nk){\rm Im} \, k_{2}(nk)  +3  (X^{-3,2}_{k})^2 (kn-2\omega) {\rm Im} \, k_2(k n-2\omega) \right\}=0 \,.
\label{secLuc10}
\end{multline}
We notice, from \eqref{secLuc11}, that $x{\rm Im}\, k_2(x)< 0$ $\forall x \neq 0$ and hence \eqref{secLuc10} holds if and only each term of the sum is zero. The Hansen coefficients have the following properties:
$\forall k \neq 0$, $X^{-3,0}_k(e)= 0 \, \Longleftrightarrow \, e=0$ and $\forall k \neq 2$, $X^{-3,2}_k(e)= 0 \, 
\Longleftrightarrow \, e=0$. This implies that $e=0$ is a necessary condition for the existence of an equilibrium.

\begin{figure}[hbt!]
\centering
\includegraphics[scale=0.9]{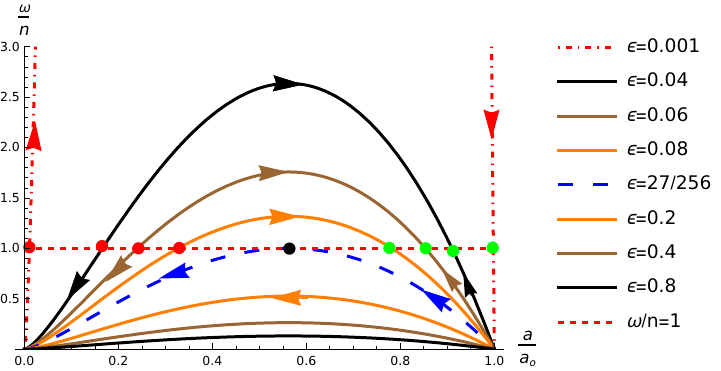}
\caption{Orbits of the equation (\ref{av3}) on the invariant plane $e=0$. The orbits are labelled by the
  total angular momentum $\ell_{\scriptscriptstyle T}$ by means of the nondimensional parameter
  $\epsilon^{-1}=\frac{\ell_{\scriptscriptstyle T}^4}{\Io \mu c^2}$. The equilibria are on the horizontal line
$\frac{\omega}{n}=1$: the green dots  represent  stable equilibria and the red dots represent  unstable
equilibria. The black dot at $\tilde a=\frac{9}{16}$, $\frac{\omega}{n}=1$ represents the
single equilibrium that occurs
for the special value $\epsilon=\frac{27}{256}$. For $\epsilon>\frac{27}{256}$ (small angular momentum)
all the solutions lead to a collision.}
\label{graphe}
\end{figure}

Conservation of angular momentum implies that the orbits of the vector field \eqref{av4} in the plane where $e=0$ are parameterized by angular momentum. Equation \eqref{secLuc13} shows that the representation of these orbits in the plane $(\tilde a, \frac{\omega}{n})$ is given  by the graphs
\begin{equation}
  \tilde a \mapsto \epsilon^{-1} \tilde{a}^{\frac{3}{2}}(1 - \tilde{a}^{\frac{1}{2}}) = \frac{\omega}{n}\,, \quad \text{for} \quad \epsilon \in (0, \infty)\,,
\end{equation}
as illustrated in Figure \ref{graphe}.

\begin{figure}[hptb!]
\centering
\begin{minipage}{0.3\textwidth}
\centering
\includegraphics[width=0.9\textwidth]{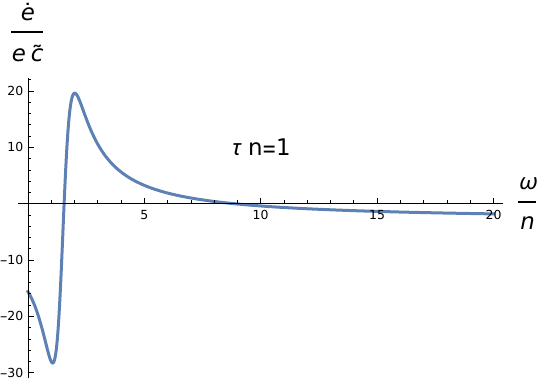}
\end{minipage}\hfill
\begin{minipage}{0.3\textwidth}
  \centering
  \includegraphics[width=0.9\textwidth]{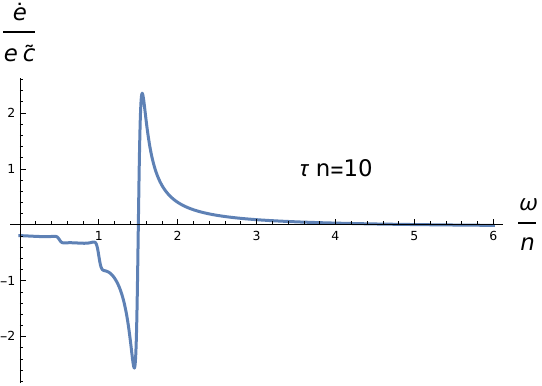}
\end{minipage}
\begin{minipage}{0.3\textwidth}
  \centering
  \includegraphics[width=0.9\textwidth]{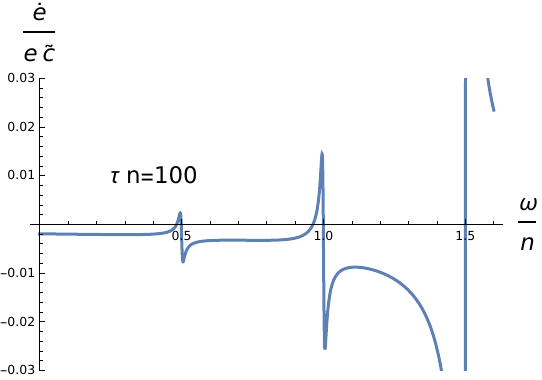}
\end{minipage}\hfill
\caption{
  The graph of  $\frac{\dot e}{e\tilde c}$ as a function
  of $\frac{\omega}{n}$ for  values of $\tau n$ equal 1, 10, and 100. For $n=10$,
  $\frac{\dot e}{e\tilde c}$ has a zero, not easily seen in the Figure, at $\frac{\omega}{n}=5.26$.
 }
\label{edot}
\end{figure}

In the significant case where $\epsilon \approx 0$, equation \eqref{secLuc13} suggests that
$0 \approx \epsilon \frac{\omega}{n} = \tilde{a}^{\frac{3}{2}}(1 - \tilde{a}^{\frac{1}{2}} \sqrt{1 - e^2})$. Up to first order in eccentricity, we have $\tilde a = 1$ and $n = n_\circ$. Using this approximation, the function $\frac{\dot e}{e}$ in equation \eqref{av4} is expressed as
\begin{equation}\begin{split}
  \frac{\dot e}{e} &= -\tilde c
  \left(\frac{6}{n^2 \tau^2 + 1} + \frac{8 \frac{\omega}{n} - 8}{4 \tau^2 n^2
    (1 - \frac{\omega}{n})^2 + 1}\right.\\ &\quad\quad\left.
  +\frac{2 \frac{\omega}{n} - 1}{\tau^2 n^2 (1 - 2 \frac{\omega}{n})^2 + 1}
  +\frac{49 (3 - 2 \frac{\omega}{n})}{\tau^2 n^2 (3 - 2 \frac{\omega}{n})^2 + 1}\right)\,,
\end{split}\end{equation}
where $n = n_\circ =$ constant and $\tilde c$ is a positive, although small, constant.
The graph of $\frac{\dot e}{e\tilde c}$ as a function of $\frac{\omega}{n}$ for various values of $\tau n$ is depicted in Figure \ref{edot}. This figure illustrates that $\frac{\dot e}{e}$ changes sign near the plane $e = 0$. Consequently, a solution with an initial eccentricity close to zero, yet sufficiently distant from the stable equilibrium at $\frac{\omega}{n} = 1$, may experience an increase in eccentricity.

The next step in understanding the dynamics of equation (\ref{av3}) involves linearization about the equilibria. It is evident from Figure \ref{graphe} that the equilibria can be parameterized by their $\tilde a$ coordinate. Thus, an equilibrium is represented by $(\omega, \tilde a) = (\omega_e, \tilde a_e)$, where, according to equation \eqref{secLuc13}, $\tilde a_e$ is the solution to
\begin{equation}
 \epsilon = \tilde{a}^{\frac{3}{2}}(1 - \tilde{a}^{\frac{1}{2}})  \,.
\end{equation}
The special equilibrium $\tilde a_e = \frac{9}{16}$, corresponding to the bifurcation value $\epsilon = \frac{27}{256}$, marked by the black dot in Figure \ref{graphe}, represents a threshold of stability: an equilibrium with $\tilde a_e < \frac{9}{16}$ is unstable, while an equilibrium with $\tilde a_e > \frac{9}{16}$ is stable.

Given that $0 < \epsilon < \frac{27}{256} \approx 0.1$, a perturbative calculation reveals that the largest root of this equation (stable equilibrium) satisfies
\begin{equation}
  \tilde a_e = 1 - 2 \epsilon - 5 \epsilon^2 + \Oc\big(\epsilon^3\big)\,.\label{sols}
\end{equation}
This approximation remains accurate up to $\epsilon = 0.05$.

At equilibrium, the orbit is circular. If $\ell_e = \ell_T - \Io n_e$ denotes the orbital angular momentum at equilibrium, then $a_e = \frac{\ell_e^2}{\mu c}$. Since $a_e = \tilde a_e a_\circ$ and $a_\circ = \frac{\ell_{\scriptscriptstyle T}^2}{\mu c}$, we obtain
\begin{equation}
  \tilde a_e = \left(\frac{\ell_e}{\ell_{\scriptscriptstyle T}}\right)^2 = \left(1 - \frac{\Io n_e}{\ell_{\scriptscriptstyle T}}\right)^2\,.
\end{equation}
Thus, $\tilde a_e$ represents the square of the ratio of orbital angular momentum to total angular momentum at equilibrium.
For the Mercury-Sun system, where $m_0$ is the mass of the Sun, $\epsilon = 6.8 \times 10^{-10}$ and $\tilde a_e \approx 1$. For the Earth-Moon system, where $m_0$ is the mass of the Moon, $\epsilon = 0.0036$ and $\tilde a_e = 0.993$. It appears that in most problems of interest, $\tilde a_e \approx 1$.

The linearization of equation (\ref{av3}) at $(\omega, \tilde a, e) = (\omega_e, \tilde a_e, 0)$ is derived
easily from equation (\ref{av4}):
\begin{equation}
  \begin{split}
    \dot \delta_\omega &= -k_\circ\frac{N \tilde n_e^2\tau}{\tilde a_e^3}\ 
    \left(2 \delta_\omega+3\frac{\tilde n_e}{\tilde a_e}\delta_a\right) =
     \left( \frac {k_\circ N \tilde n_e\tau}{\tilde a_e^3}\right)\ \tilde n_e
  \left(-2 \delta_\omega-3\frac{\tilde n_e}{\tilde a_e}\delta_a\right),\\ 
  \dot \delta_a &=  k_\circ\frac{2 N \tilde n_e \epsilon \tau}{\tilde a_e^4}\
  \left(2 \delta_\omega+3\frac{\tilde n_e}{\tilde a_e}\delta_a\right) = \left( \frac {k_\circ N \tilde n_e\tau}{\tilde a_e^3}\right)
  \frac{2  \epsilon }{\tilde a_e}\
  \left(2 \delta_\omega+3\frac{\tilde n_e}{\tilde a_e}\delta_a\right),\\ 
  \dot e &= -k_\circ \frac{N \tilde n_e^2\epsilon\,\tau} { \tilde a_e^{5}}\
    \frac{7}{1+\tilde n_e^2\tau^2} e = -\left( \frac {k_\circ N \tilde n_e\tau}{\tilde a_e^3}\right) \frac{ \tilde n_e\epsilon} { \tilde a_e^{2}}\
    \frac{7}{1+\tilde n_e^2\tau^2} e,
\end{split}
\label{av5}
\end{equation}
where $\tilde n_e = n_0\,\frac{1}{\tilde a_e^{3/2}}$.
Each equilibrium has: one eigenvalue equal to zero, associated with the conservation of angular momentum; one negative eigenvalue $\lambda_e = -\frac{7 \tilde n_e \epsilon}{\tilde a_e^2(\tilde n_e^2 \tau^2 + 1)}\left(\frac{k_\circ N \tilde n_e\tau}{\tilde a_e^3}\right)$, with an eigenvector tangent to the eccentricity axis; and one eigenvalue $\lambda_0 = -\frac{2 \tilde n_e (\tilde a_e^2 - 3 \epsilon)}{\tilde a_e^2}\left(\frac{k_\circ N \tilde n_e\tau}{\tilde a_e^3}\right)$, with an eigenvector in the plane $e=0$ and tangent to the surface of constant angular momentum. As expected, $\lambda_0 = 0$ in the critical case where $\epsilon = \frac{27}{256}$ and $\tilde a_e = \frac{9}{16}$, $\lambda_0 > 0$ if $\tilde a_e < \frac{9}{16}$, and $\lambda_0 < 0$ if $\tilde a_e > \frac{9}{16}$.

\begin{figure}[hptb!]
\centering
\begin{minipage}{0.5\textwidth}
\centering
\includegraphics[width=0.9\textwidth]{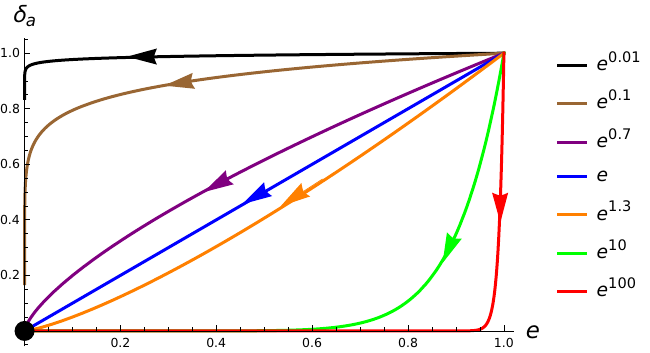}
\end{minipage}\hfill
\begin{minipage}{0.45\textwidth}
  \centering
  \includegraphics[width=0.9\textwidth]{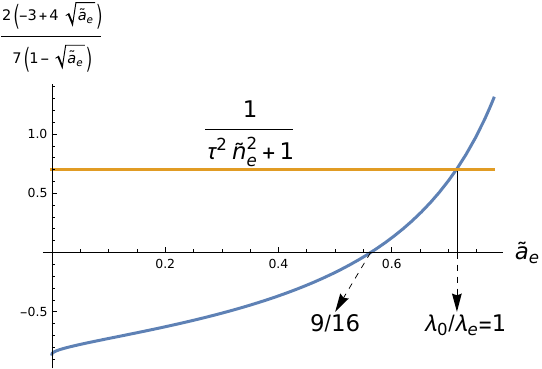}
\end{minipage}\hfill
\caption{LEFT: The figure shows possible orbits on the eccentricity-semi-major axis ($\delta_a = \tilde a - \tilde a_e$) plane, $\delta_a = \text{constant}\ e^{\lambda_0/\lambda_e}$ with $\text{constant} = 1$ for various $\lambda_0/\lambda_e$ values. RIGHT: A graphical method to find the special value of $\tilde a_e$, where $\lambda_0/\lambda_e = 1$, as a function of $\tau \tilde n_e$.}
\label{figae}
\end{figure}

Consider a solution to equation (\ref{av4}) that satisfies $\lim_{t \to \infty}\big(e(t), \tilde a(t)\big) = \big(0, \tilde a_e\big)$, and let $\delta_a(t) = \tilde a(t) - \tilde a_e$. At a certain time $\tilde t$, this solution is sufficiently close to $(0,\tilde a_e)$ for the linear approximation to be valid. Since $\delta_a(t) = e^{\lambda_0 (t - \tilde t)}\delta_a(\tilde t)$ and $e(t) = e^{\lambda_e (t - \tilde t)}e(\tilde t)$, we conclude that near the equilibrium,
\begin{equation}
  \delta_a(e) = \tilde a(e) - \tilde a_e = \underbrace{\frac{\delta_a(\tilde t)}{e^{\lambda_0/\lambda_e}(\tilde t)}}_{=\text{constant}} e^{\lambda_0/\lambda_e}\,,\label{ae}
\end{equation}
where
\begin{equation}
  \frac{\lambda_0}{\lambda_e} = \frac{2 \left(\tilde a_e^2 - 3 \epsilon \right)}{7 \epsilon} \left(\tilde n_e^2 \tau^2 + 1\right) = \frac{8 \left(\sqrt{\tilde a_e} - \frac{3}{4}\right)}{7 \left(1 - \sqrt{\tilde a_e}\right)} \left(\tilde n_e^2 \tau^2 + 1\right)\,.
\end{equation}
Regardless of the value of the constant factor in equation \eqref{ae}, which in Figure \ref{figae} we assume to be one, the orbit's geometry near the equilibrium is controlled by the ratio $\frac{\lambda_0}{\lambda_e}$. In Figure \ref{figae} LEFT, we illustrate how the orbit changes as $\frac{\lambda_0}{\lambda_e}$ varies, with the ratio $\frac{\lambda_0}{\lambda_e} = 1$ being a critical value. For $\frac{\lambda_0}{\lambda_e} > 1$, the orbit approaches the equilibrium along the $e$-axis, and for $0 < \frac{\lambda_0}{\lambda_e} < 1$, the orbit approaches the equilibrium along the $\delta_a$ axis. In Figure \ref{figae} RIGHT, we demonstrate how to determine the special value of $\tilde a_e$, corresponding to $\frac{\lambda_0}{\lambda_e} = 1$, as a function of the parameter $\tau \tilde n_e$. The maximal value of this special $\tilde a_e$ is $\frac{169}{225} \approx 0.75$, achieved when $\tau = 0$.

It appears that in most problems of interest, $\epsilon$ is very small, $\tilde a_e \approx 1$, and $\lambda_0/\lambda_e \gg 1$, indicating that solutions approach the stable equilibrium along the $e$-axis, namely the weak-stable manifold of the equilibrium.

\subsection{Slow-fast systems and singular perturbation theory} 

For $\epsilon \approx 0$
equation \eqref{av4} has the form of a slow-fast system:
\begin{equation}
\begin{split}
\dot{x} & = f(x,y,\epsilon),\\ 
\dot{y} & = \epsilon g(x,y,\epsilon),
\end{split}
\label{secLuc8}
\end{equation}
with $x=\omega\in \R$ as the fast variable and $y=(e,\tilde{a})\in \R^2$ as the slow variables \cite{fenichel1979geometric}.

Given an initial condition in the state space $\big\{\omega,e,\tilde{a}\big\}$,
the value of $\omega$ varies while $(e,\tilde{a})$ stays nearly constant
until the state reaches the slow manifold
\begin{equation}
\Sigma_s(0):=\left\{\dot \omega(\omega,e,\tilde{a})=0\right\}=\left\{ {\cal A}_0(\omega,e,\tilde{a})=0\right\}\quad\text{(s denotes slow)}\,,
\label{secLuc12}
\end{equation}
where ${\cal A}_0$ is given in equation (\ref{av}).

When $(x,y_0)$ 
  is not close to $\Sigma_s(0)$, the fast dynamics is governed by the layer problem, $\dot{x}=f(x,y_0,0)$. 
Here, the fast dynamics corresponds to the fast spin variation with fixed $e$ and $\tilde{a}$. The spin decreases on points above  $\Sigma_s(0)$ and decreases on points under
  $\Sigma_s(0)$, see Figure \ref{retratomedio}. Close to the slow manifold $\Sigma_s(0)$, the dynamics is approximated by the reduced problem, where the fast variable is given by an implicit function, solution of $f(\Phi(y),y,0)=0$, and the slow variable solves
the differential equation on $\Sigma_s(0)$, $\dot{y} = g(\Phi(y),y,0)$. The implicit function theorem ensures that $\Phi$ is locally determined at $(x_0,y_0) \in \Sigma_s$ if  $\partial_x f(x_0,y_0,0)\neq 0$. In this case, $\Sigma_s(0)$ is called normally hyperbolic at $(x_0,y_0)$. The results from geometric singular perturbation theory \cite{fenichel1979geometric} state that if the system \eqref{secLuc8} has a normally hyperbolic slow manifold $S_0$, for each small $\epsilon>0$ exists an invariant manifold $S_\epsilon$ diffeomorphic to $S_0$ which is stable (unstable) if $\partial_x f<0 $ ($\partial_x f>0 $) on $S_0$. We will denote by $\Sigma_s(\epsilon)$ the union of the hyperbolic components of perturbed slow manifold in \eqref{secLuc12}.

\begin{figure}[hbt!]
\centering
\includegraphics[scale=0.34]{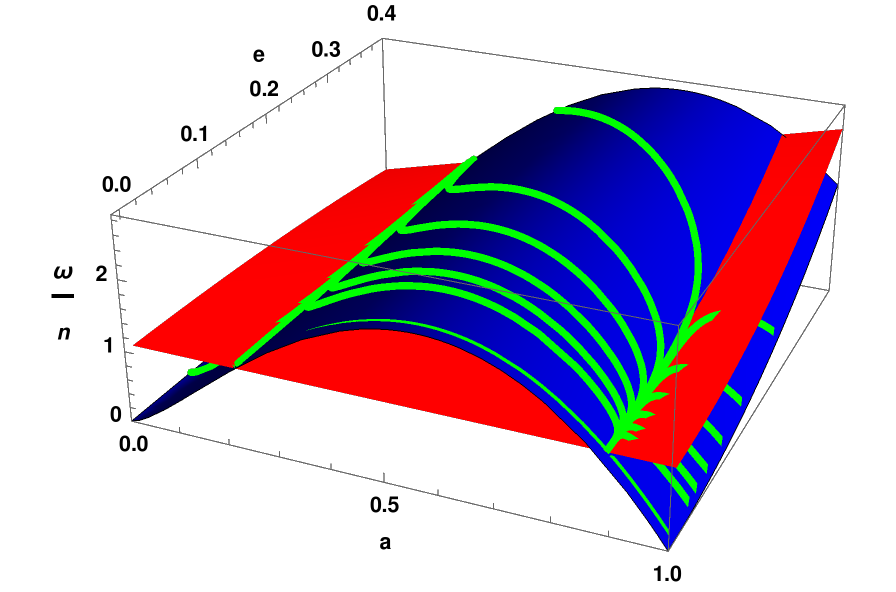}
\includegraphics[scale=0.41]{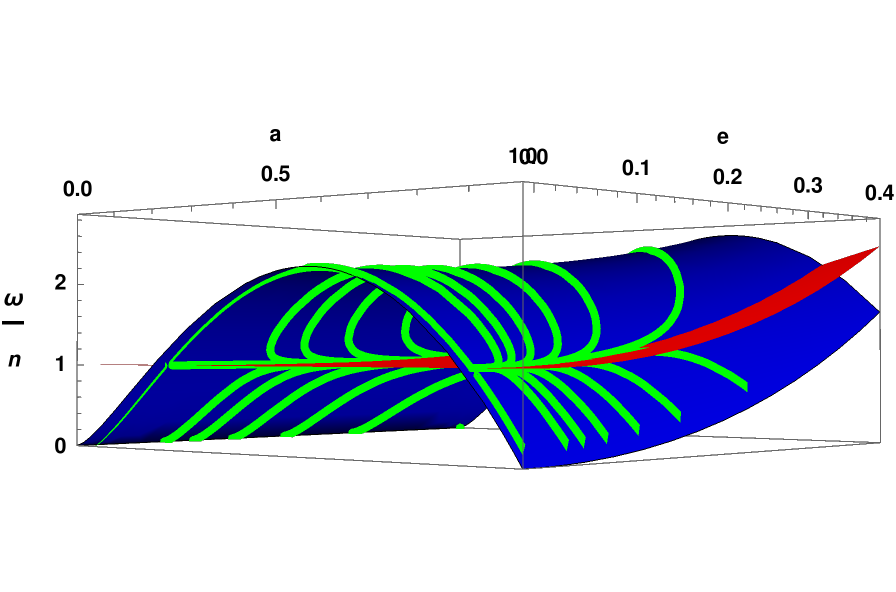}
\caption{The phase space close to the synchronous states $\omega/n=1$, $e=0$. The blue
  surface represents a level set of the  angular momentum and the red surface represents the
  slow manifold $\Sigma_s(\epsilon)$. Both surfaces and the  plane $e=0$ intersect only at the equilibria.
  The stable separatrix of the saddle point delimits the basin of attraction of the node and the region whose solutions tend to the collision $a=0$.}
\label{bifurcacao}
\end{figure}

The dynamics across the entire phase space can be elucidated by examining the geometry of the slow manifold \eqref{secLuc12}. Within the first octant ${\cal B}_1 := \{\omega > 0, e > 0, a > 0\}$, $\Sigma_s(0)$ possesses a single connected component that splits ${\cal B}_1$ into two regions. The conservation of angular momentum reduces the
analysis to a two-dimensional problem. A diagram illustrating the local behavior of orbits near the stable equilibrium is presented in Figure \ref{figae} LEFT. A global illustration of the flow on a level set of angular momentum is shown in Figure \ref{bifurcacao}.

\section{Spin-Orbit Resonances}
\label{secso}

In this section, we assume that the ratio $\frac{\omega}{n}$ is at most on the order of tens, so that
\begin{equation}
  \left|\epsilon \frac{\omega}{n}\right| \ll 1.
\end{equation}

Under this condition, equation \eqref{secLuc13}, i.e., 
$ \tilde{a}^{\frac{3}{2}}(1 - \tilde{a}^{\frac{1}{2}} \sqrt{1 - e^2}) = \epsilon \frac{\omega}{n}$,
yields two solutions for $\tilde a$. The first solution is $\tilde a = \left(\epsilon \frac{\omega}{n}\right)^{2/3} + \Oc(\epsilon)$. This solution closely approximates the surface of constant angular momentum in a region that includes the unstable equilibrium $\tilde a_e \approx 0$. This approximation is depicted in Figure \ref{graphe} by the nearly vertical red dot-dashed line near $\tilde a_e \approx 0$.  We will not focus on this region. The second solution is $\tilde a = \frac{1}{1 - e^2} + \Oc(\epsilon)$, which is of primary interest. This solution approximates the surface of constant angular momentum in a region containing the stable equilibrium $\tilde a_e \approx 1$. This approximation is depicted in Figure \ref{graphe} by the nearly vertical red dot-dashed line near $\tilde a_e \approx 1$.  Disregarding the error of order $\epsilon$, we have $\tilde a_e = 1$, $a_e = a_\circ$, and 
\begin{equation}
    \tilde a = \frac{1}{1 - e^2} \Rightarrow a = a_\circ \frac{1}{1 - e^2}, \quad
    \text{where} \quad a_\circ = \frac{\ell_{\scriptscriptstyle T}^2}{\mu c}.
\label{secLuc133}
\end{equation}
In the subsequent analysis we use these approximations.

\begin{figure}[hptb!]
\centering
\begin{minipage}{0.3\textwidth}
\centering
\includegraphics[width=0.9\textwidth]{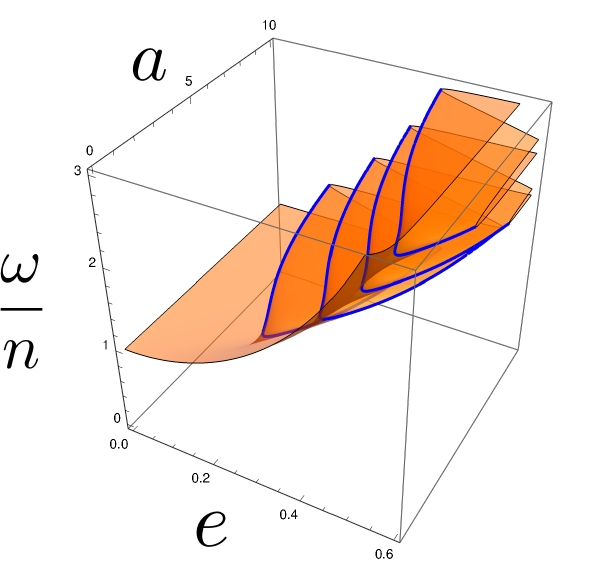}
\end{minipage}\hfill
\begin{minipage}{0.3\textwidth}
  \centering
  \includegraphics[width=0.9\textwidth]{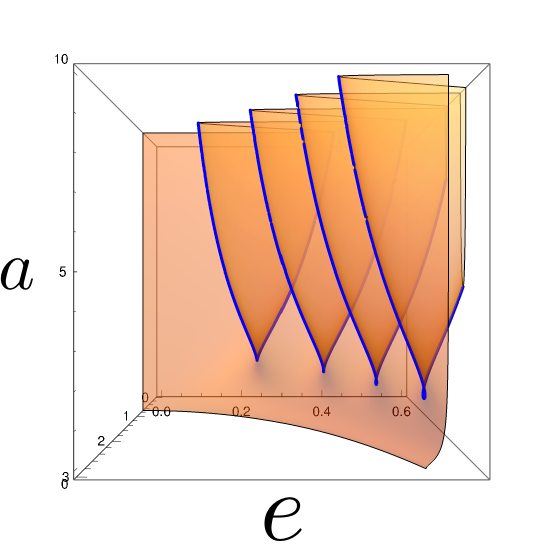}
\end{minipage}
\begin{minipage}{0.3\textwidth}
  \centering
  \includegraphics[width=0.9\textwidth]{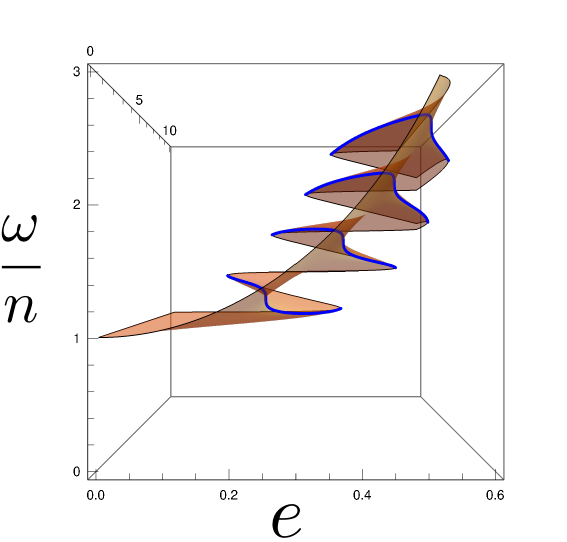}
\end{minipage}\hfill
\caption{
  Figure showing three views of the slow manifold $\Sigma_s(0)$, which is the orange surface, and
  the fold  curves   in blue.
 }
\label{fold}
\end{figure}

The geometry of the slow manifold $\Sigma_s(0)$ plays a crucial role in the capture into spin-orbit resonance, particularly  where $\Sigma_s(0)$ is not normally hyperbolic. The slow manifold becomes non-normally hyperbolic at points where the projection map from $\Sigma_s(0)$ to the $\{a,e\}$ plane is singular. These generic singular points of the projection are known as folds and collectively form the ``fold curves''. In Figure \ref{fold}, the fold curves are depicted in blue on the slow manifold $\Sigma_s(0)$, which is represented as an orange surface.
Although the fold curves themselves are smooth, their projection onto the $\{a,e\}$ plane includes
singular points termed ``cusps'', at which a moving point on the projection reverses direction.
A cusp point on a fold curve occurs where the tangent to the curve becomes parallel to the $\omega$-axis.
The flow dynamics near a fold are extensively described in the literature \cite{krupa2001extending}.

We illustrate the so called phenomenon of capture into spin-orbit resonance by a concrete example presented in \cite{correia2014deformation} and \cite{correia2018effects}. We use the parameters of the exoplanet HD80606b and its hosting star, namely $m_0 = 2008.9 \cdot 10^{30}$kg, $m = 7.746 \cdot 10^{28}$kg, $\Io = 8.1527 \cdot 10^{40}$kg ${\rm m^2}$. The initial conditions are chosen as $a=0.455$au, $e=0.9330$ and $\omega=4\pi \,  {\rm rad}/{\rm day}$ and hence $\epsilon = 1.35 \cdot 10^{-8}$. The parameters of the rheology are $k_\circ = 0.5$ and $\tau = 10^{-2}$year.

\begin{figure}[hbt!]
\centering
\includegraphics[scale=0.41]{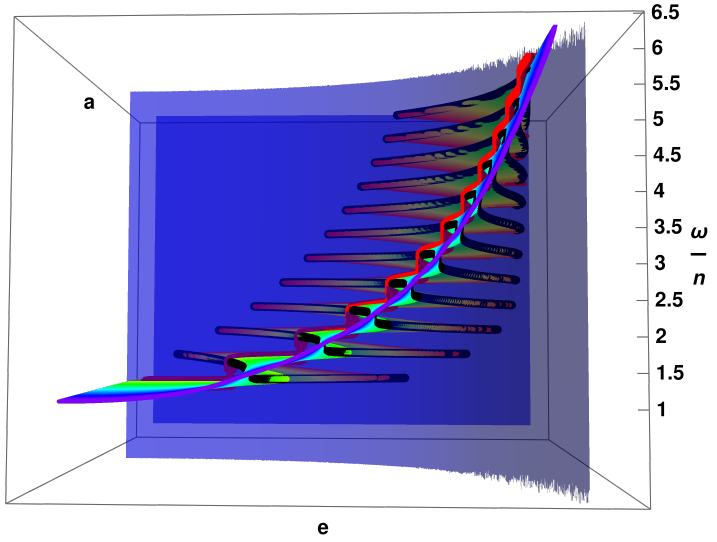}
\includegraphics[scale=0.41]{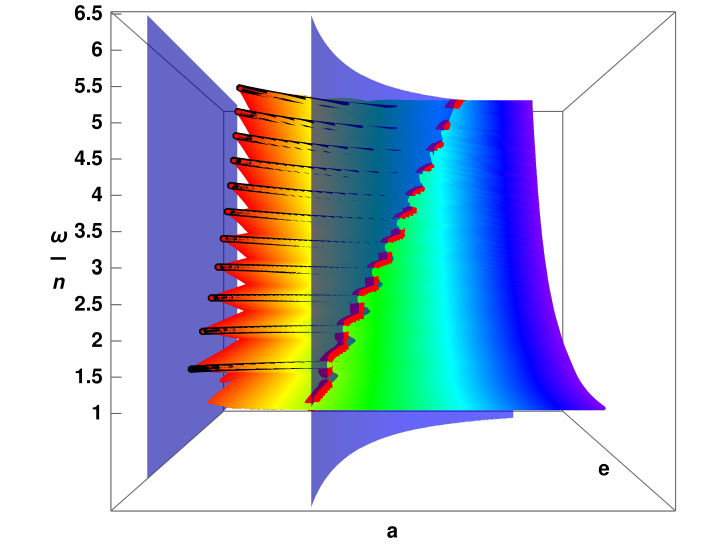}
\includegraphics[scale=0.41]{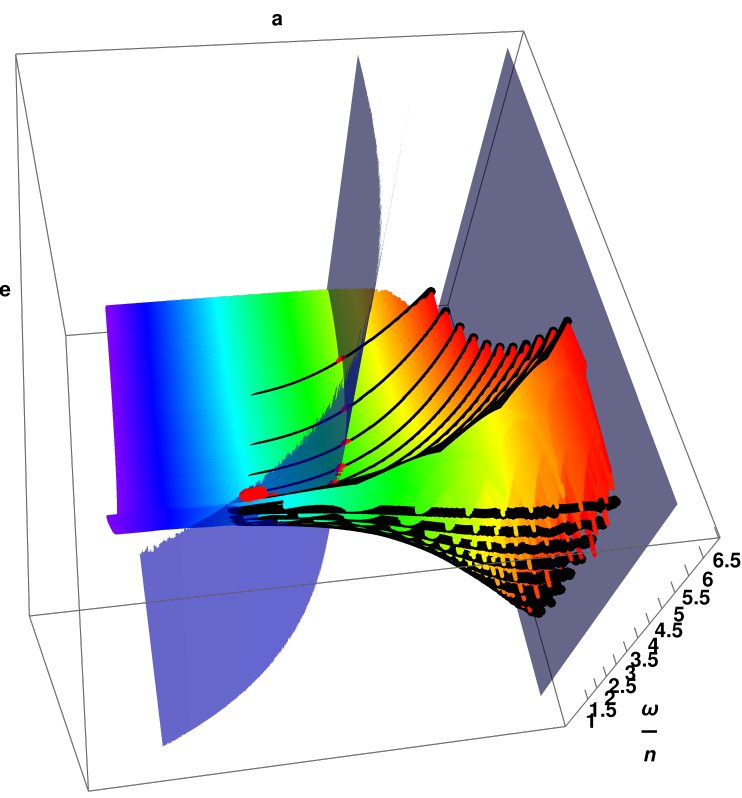}
\includegraphics[scale=0.23]{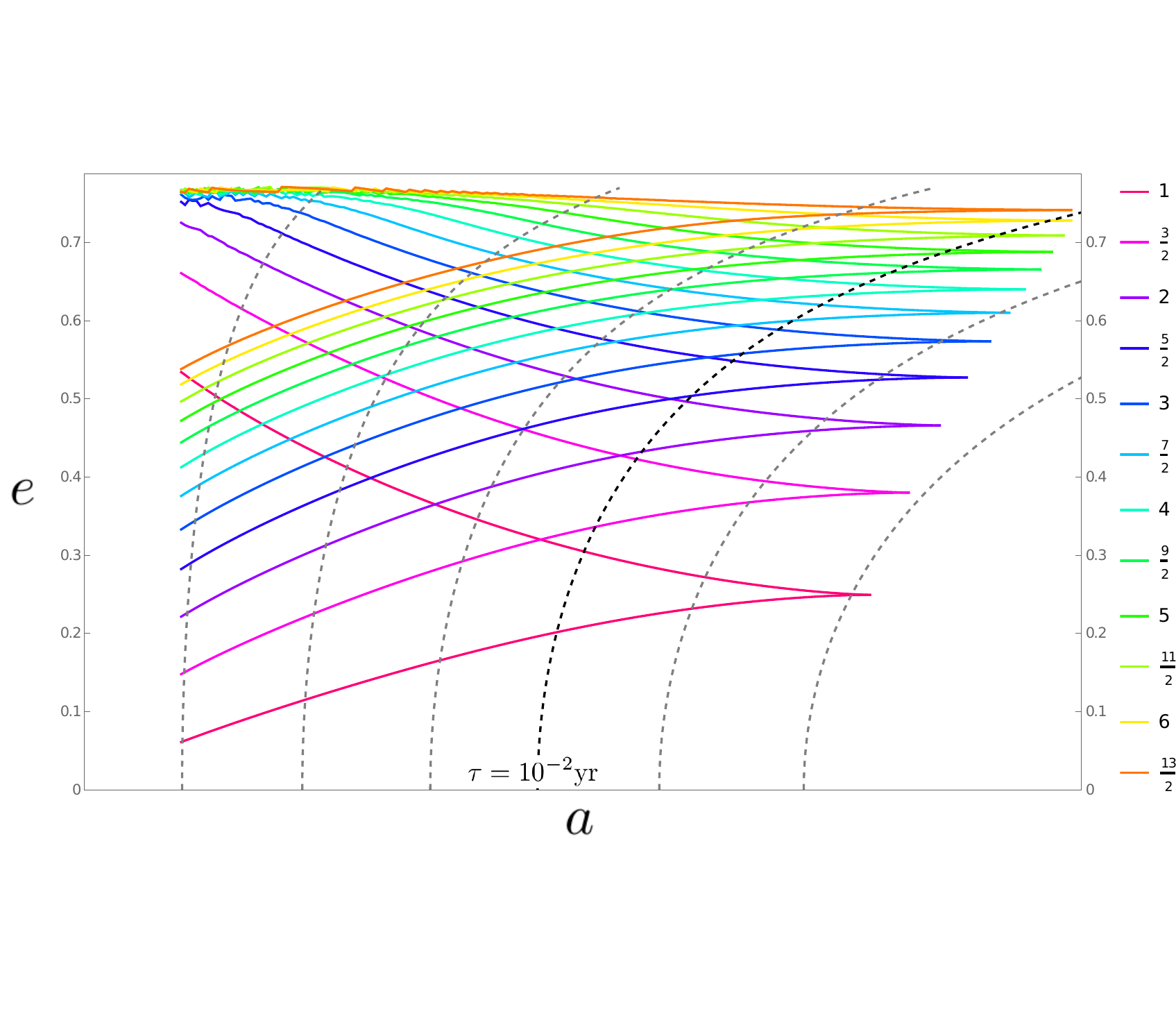}
\caption{Geometrical perspective of capture into spin-orbit resonances.  The slow manifold $\Sigma_s(0)$
  loses normal hyperbolicity at fold curves (black),
  characterized by $\partial_{\tilde{\omega}} {\cal A}_0 = 0$. The fold curves become parallel to the
  $\frac{\omega}{n}$-axis at the cusp points.
  The blue surface represents a level set of angular momentum, as depicted in Figure \ref{bifurcacao}. The red
  curve represents a solution of the complete system, which exhibits jumps when crossing the fold curves. In the lower-right frame, the projection of the cusp-shaped curves onto the $(a,e)$ plane is displayed. Each curve is annotated with an integer or half-integer, symbolizing a resonance $\omega/n = \frac{k}{2}$, for $k = 1, \ldots, 13$, as noted on the right side of the figure. The dashed lines correspond to projections of the constant angular momentum surfaces $a = \frac{\ell_{\scriptscriptstyle T}^2}{\mu c}\frac{1}{1 - e^2}$. If the total angular momentum $\ell_{\scriptscriptstyle T}$ is sufficiently large such that the curve $a = \frac{\ell_{\scriptscriptstyle T}^2}{\mu c}\frac{1}{1 - e^2}$ does not intersect the projection of the fold curve associated with a specific $\omega/n = \frac{k}{2}$ spin-orbit resonance, then the $k:2$ resonance is precluded for that angular momentum value.}
\label{exoplaneta}
\end{figure}

In  Figures  \ref{exoplaneta} (top panels),  the red curve represents  a trajectory of 
the fundamental equations, given in Section \ref{eqsec},  which
was obtained by means of numerical integration.  The numerically computed trajectory
has  consecutive transitions between stable branches of the perturbed slow manifold $\Sigma_s (\epsilon)$. 
This trajectory shows a slow decrease of the eccentricity towards $e=0$ while the spin-orbit ratio has fast transitions between integers and half-integers  with final value $\omega/n=1$. The stable branches
of $\Sigma_s(\epsilon)$ are quite flat (parallel to the $(e,a)$-plane) near the planes $\frac{\omega}{n}=\frac{k}{2}$,
$k\in \Z$. These results are detailed in Figure 14 from \cite{correia2018effects}. We can
observe in Figure \ref{exoplaneta} the full agreement between the solution of the fundamental equations
and the fast-slow-geometric analysis of the averaged equations. 

The projection of the fold curves to the plane $(a,e)$ are shown in Figure \ref{exoplaneta} DOWN-RIGHT.
Each curve contains  a cusp singularity and is labeled by an integer or  half-integer.
 A point initially over $(a,e)$ can be attracted to a resonance
 $\frac{\omega}{n}=\frac{k}{2}$ only if it is inside a dashed curve that intersects the curve labeled by $\frac{k}{2}$; see caption of Figure
 \ref{exoplaneta} for further information.

\subsection{Spin-Orbit Resonances Requires Large Relaxation Times $\tau$.}

The approximation $\tilde a=(1-e^2)^{-1}$ and equation \eqref{ncirc2} imply $n=n_\circ(1-e^2)^{3/2}$. 
The imaginary part of the  Love number \eqref{secLuc11}
can then be written as
\begin{equation}\begin{split}
    {\rm Im} k_2(k n-2\omega) &=
    k_\circ  \frac{\, 2 \tau n (\omega/n - k/2)}{1+ (2\tau n)^2 (\omega/n - k/2)^2} \\ & = \sqrt{\tilde{\epsilon}} k_\circ (1-e^2)^{\frac{3}{2}}  \frac{\, (\omega/n - k/2)}{\tilde{\epsilon} + (1-e^2)^3(\omega/n - k/2)^2} \,,\end{split}
\label{secLuc15}
\end{equation}
where
\begin{equation}
  \tilde{\epsilon} := \frac{1}{(2\tau n_\circ)^2}\,,
  \qquad  n_\circ=\frac{c^2 \mu}{\ell_{\scriptscriptstyle T}^3}\,.
\label{secLuc16}
\end{equation}

For $e=0$, the slow manifold lacks any fold points for any value of $\tau > 0$, as illustrated in Figure \ref{graphe}. Consider a fixed value $e_1 > 0$ for $e$. Equations (\ref{V}) and \eqref{secLuc15}
imply the existence of at least $j-1$ fold points in the region $\{0 < e < e_1, 0 < \frac{\omega}{n} < C\}$, where $C > 0$ represents a positive constant, if and only if
\begin{equation}
\begin{split}
   \frac{\omega}{n} \mapsto &
   \Big(X^{-3,2}_{2}(e_1)\Big)^2  \frac{\, (\omega/n - 1)}{\tilde{\epsilon} + (1-e_1^2)^3(\omega/n - 1)^2} \\
   & +
\sum\limits_{k\ne 2}
 \Big(X^{-3,2}_{k}(e_1)\Big)^2  \frac{\, (\omega/n - k/2)}{\tilde{\epsilon} + (1-e_1^2)^3(\omega/n - k/2)^2}
\end{split}
\label{V2}
\end{equation}
has $j$ zeroes for $\frac{\omega}{n} \in (0, C)$.

Given that $X^{-3,2}_{k}(0) = 1$ and $X^{-3,2}_{k}(e_1) = \Oc(e_1)$, function \eqref{V2} can be expressed as
$ \frac{\, (\omega/n - 1)}{\tilde{\epsilon} + (\omega/n - 1)^2} + \Oc(e_1^2)$. For $0 < \frac{\omega}{n} < C$ and a fixed $\tilde \epsilon > 0$, this function exhibits a single zero near $\frac{\omega}{n} = 1$ if $e_1 > 0$ is sufficiently small. Furthermore, for a fixed $e_1 > 0$ and $\tilde \epsilon = 0$, function \eqref{V2} presents poles for every $\frac{\omega}{n} = \frac{k}{2}$, $k \in \mathbb{Z}$, thereby ensuring at least one zero in each interval $(k, k + \frac{1}{2})$, where $k$ is any half-integer. A continuity argument suggests that if $\tilde{\epsilon}$ is sufficiently close to zero (implying $\tau$ is sufficiently large), then for any fixed $e_1$, function \eqref{V2} will have zeroes near $j/2$, for $j = 1, 2, \ldots$. This analysis indicates that, particularly for small $e_1 > 0$, the condition $\tilde \epsilon \ll 1$ (equivalently, $\tau\gg 1$) is a necessary condition for the creation of folds in the slow manifold $\Sigma_s(0)$.

For the  Earth-Moon system, 
where $m_0$ is the mass of the Moon, $\epsilon=0.0036$ and $n_\circ^{-1}=7.6$ days,  a value $\tau>76$ days
gives $\tilde \epsilon<0.0025$. For the  Mercury-Sun system, where $m_0$ is the mass of the Sun,
$\epsilon=6.8\times 10^{-10}$ and $n_\circ^{-1}=13$ days, a value   $\tau>130$ days gives
$\tilde \epsilon<0.0025$.
In the case of the parameters chosen for HD80606b, $\tilde{\epsilon} \approx 1.28 \cdot 10^{-5} $.

For $\tilde \epsilon\ll 1$ and close to a resonance $\omega/n = j/2$, $j\in\{2,3,\ldots\}$,
$\Sigma_s(0)$ can be approximately computed  as a power series in $\epsilon$. 
If we substitute
\[
  \omega/n = j/2 +  \Phi(j/2,e,\tilde{\epsilon}) = j/2 + \Phi_1(j/2,e)\tilde{\epsilon} +
  \Phi_2(j/2,e)\tilde{\epsilon}^2 + \ldots\]
into  the equation
\begin{equation}
 \sum\limits_{k=-\infty}^\infty
\Big(X^{-3,2}_{k}(e)\Big)^2  \frac{\, (\Phi(j/2,e,\tilde{\epsilon}) +(j- k)/2)}{\tilde{\epsilon} + (1-e^2)^3(\Phi(m/2,e,\tilde{\epsilon}) +(j- k)/2)^2}=0,
\end{equation}
and solve the resulting equation for the coefficient of  $\tilde \epsilon$ and $\tilde \epsilon^2$ we obtain, 
\begin{equation}
\Phi_1(j/2,e) =  \frac{2}{(1-e^2)^3(X_{j}^{-3,2})^2} \sum_{k=1}^{+\infty}  \frac{(X_{j+k}^{-3,2})^2 - (X_{j-k}^{-3,2})^2}{k},
\end{equation}
\begin{multline}
\Phi_2(j/2,e) =  \Phi_0(j/2,e)^3 (1-e^2)^3 \\ +\frac{4}{(1-e^2)^3(X_{j}^{-3,2})^2}\left(\Phi_0(j/2,e) \sum_{k=1}^{+\infty}  \frac{(X_{j+k}^{-3,2})^2 + (X_{j-k}^{-3,2})^2}{k^2}  \right. \\ \left. +\frac{2}{(1-e^2)^3}\sum_{k=1}^{+\infty}  \frac{(X_{j-k}^{-3,2})^2 - (X_{j+k}^{-3,2})^2}{k^3} \right).
\end{multline}

\begin{figure}[hbt!]
\centering
\includegraphics[scale=0.53]{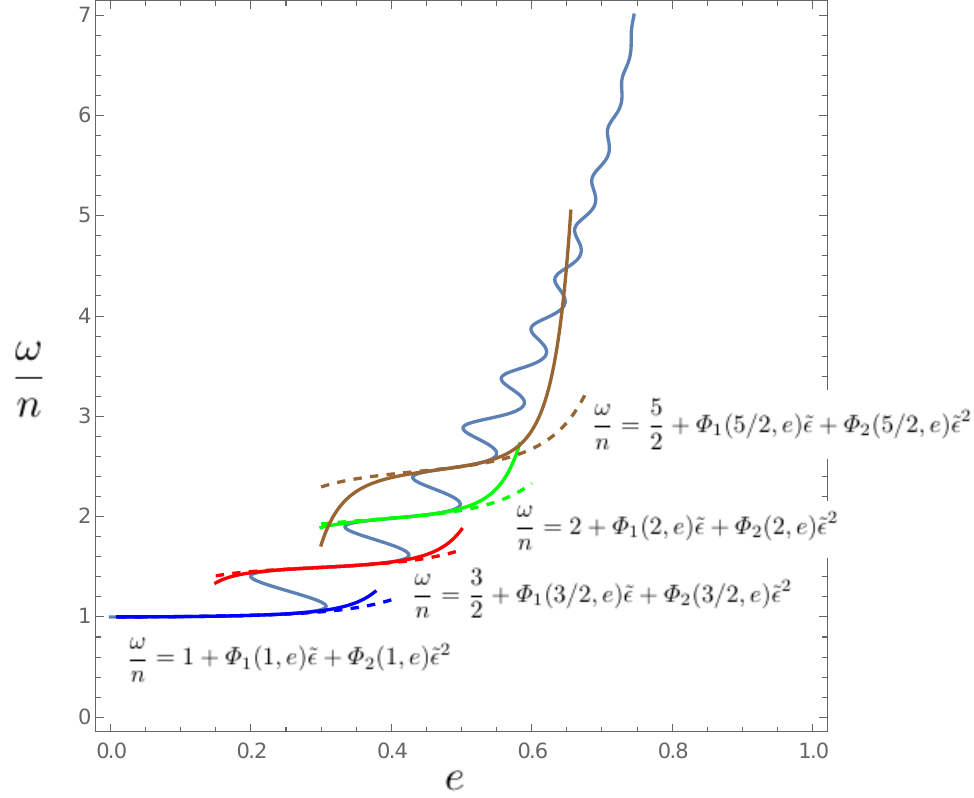}
\caption{Approximation of the resonances $\omega/n \approx j/2$, for $j=2,3,4,5$. The dashed lines correspond to the approximation up to $\mathcal{O}(\tilde{\epsilon})$ and the continuous lines up to $\mathcal{O}(\tilde{\epsilon}^2)$. In this graph we use the parameters of HD80606b, $\tilde{\epsilon} \approx 1.28 \cdot 10^{-5} $.}
\label{aprox_res}
\end{figure}

We emphasize that the functions $\Phi(j/2, e, \tilde{\epsilon})$ represent the $\mathcal{O}(\epsilon^0)$ approximations of the slow invariant manifold $\Sigma_s(\epsilon)$. These functions determine the dynamics of the reduced system, serving as the initial step in comprehending the flow on $\Sigma_s(\epsilon)$. Further exploration of this flow constitutes a subject for future work. Figure \ref{aprox_res} illustrates the approximation of $\Sigma_s(0)$ on some resonances.     

\begin{figure}[hbt!]
\centering
\includegraphics[scale=0.33]{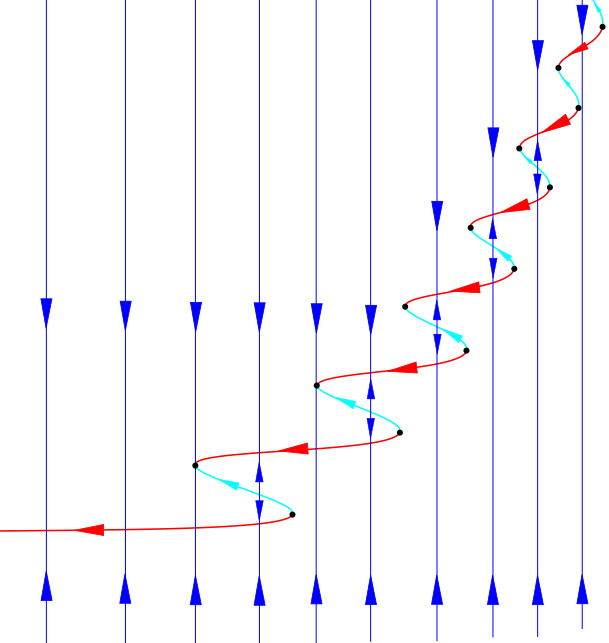}
\includegraphics[scale=0.41]{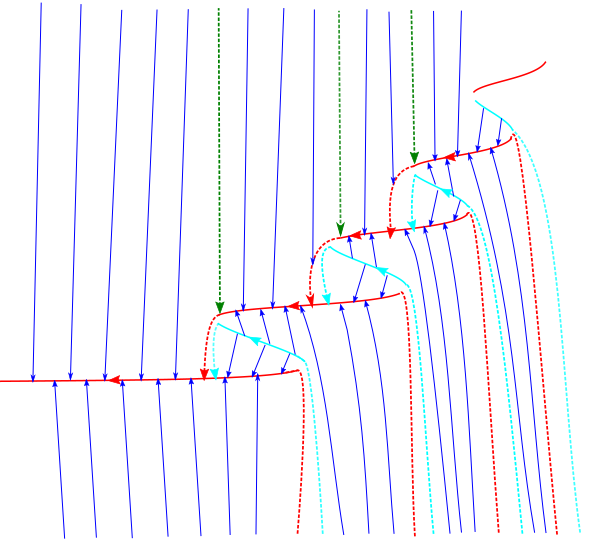}
\caption{The phase portrait of the averaged system \eqref{av3}. In the left, the lines in dark blue represent the solutions of the layer problem, the light blue represent the solutions of the reduced problem on the unstable branches of $\Sigma_s(0)$ and the red lines the solutions of the reduced problem on the stable branches. The black points are those at which the manifold ceases to be normally hyperbolic, the generic fold points. In the right, we depict the perturbed flow, i.e. for $\epsilon>0$. The solutions close to the fold points, where the jumps occur, are characterized in \cite{krupa2001extending}, see for instance Figure 2.1 on page 289. The perturbed fast flow is also represented in dark blue, except for some especial solutions. We highlight, in dark green, the solutions incident on the fold points, these solutions delimit the basin of attraction of the various spin-orbit resonances for prograde motions ($\omega>n$). In red and light blue are represented invariant manifolds that persisted under the perturbation. The continuation of these manifolds, dashed red and light blue, also delimit the portion of the resonances' basin of attraction for retrograde motions ($\omega<n$). We remark that, since the normally hyperbolic components of $\Sigma_s(0)$ are not compact, the persisting manifolds are not necessarily unique, however the qualitative behavior of the flow is the same, see \cite{krupa2001extending} for details. This geometric perspective also assists in the significant problem in tide theory concerning the probability of capture into spin-orbit resonances, such probabilities are proportional to the area of the basins of attraction.}
\label{retratomedio}
\end{figure}

We end this section with a topological description of the slow-fast dynamics of equation \eqref{av3}.
In Figure \ref{retratomedio} we present a sketch of  flow lines for $\epsilon=0$ (LEFT panel) and
$\epsilon >0$ small (RIGHT panel). Explanations are given in the Figure caption. 
The orientation of the fast flow lines  was previously examined in Section \ref{sect_fast}.
The orientation of the slow flow lines is determined by the monotonic decrease in eccentricity on $\Sigma_s(0)$.
This is a  consequence of the same argument employed to determine the equilibria,
as presented in equation  \eqref{secLuc10}.

\newpage

\section{Conclusion}

\label{conclusions}

In this paper, we presented a set of equations for the evolution of the orbital elements in the gravitational two-body problem under the influence of tides. These equations, previously obtained by other authors, were derived here through a two-step procedure. Initially, we used the fact that tidal deformations are very small to demonstrate the existence of an invariant manifold, which we have termed the deformation manifold. Although our arguments are mathematically sound, they lack the appropriate quantifiers. The second step involves averaging the equations on the deformation manifold. This step is contingent upon the first, leading to uncertainties about whether the averaged equations are mathematically coherent with the large values of $\tau n$ used in Section \ref{secso}. In the physics literature, employing large values of $\tau n$ in the averaged equations has been common practice.

Analyzing the averaged equations mathematically presents a significant challenge due to the analytical complexity of the vector field, defined by infinite sums of Hansen coefficients, which are themselves infinite series in powers of eccentricity.

Given the scientific significance of this problem, it warrants investigation from a mathematical perspective. The geometric theory of singular perturbation, potentially incorporating multiple time scales as suggested in our companion paper \cite{rr2024b}, appears to be a suitable  mathematical framework to address this challenge.

\section*{Acknowledgements}
C.R. is partially supported by by FAPESP grant 2016/25053-8. L.R.S.  is supported in part by FAPEMIG (Funda\c{c}\~ao de Amparo \`a Pesquisa no
Estado de Minas Gerais) under Grants No. RED-00133-21 and APQ-02153-23.

\section*{Conflict of interest}

On behalf of all authors, the corresponding author states that there is no conflict of interest.

\bibliographystyle{plainnat}
\bibliography{mybibliography}

\end{document}